\documentclass[oldversion]{aa}  
\usepackage{graphicx,amsmath}
\usepackage{amssymb}
\usepackage{color}
\usepackage{natbib}		
\usepackage{url}
\usepackage{multirow}
\usepackage{threeparttable}   
\bibpunct{(}{)}{;}{a}{}{,} 

\def\approxinf{%
  \def\p{%
    \setbox0=\vbox{\hbox{$<$}}%
    \ht0=0.6ex \box0 }%
  \def\s{%
    \vbox{\hbox{$\sim$}}%
  }%
  \mathrel{\raisebox{0.7ex}{%
      \mbox{$\underset{\s}{\p}$}%
    }}%
}

\begin{document}

   \title{Strong \ion{H}{i} Lyman-$\alpha$ variations from the 11\,Gyr-old host star Kepler-444: a planetary origin ?}
   				   
   \author{
   V. Bourrier\inst{1},
   D. Ehrenreich\inst{1},   
   R. Allart\inst{1},   
   A. Wyttenbach\inst{1},
   T. Semaan\inst{1},
   N. Astudillo-Defru\inst{1},
   A. Gracia-Bern\'a\inst{2},
   C. Lovis\inst{1},
   F. Pepe\inst{1},	 
   N. Thomas,\inst{2},	
   S. Udry\inst{1}}	   
   
\authorrunning{V.~Bourrier}
\titlerunning{Kepler-444 at Lyman-$\alpha$}

\offprints{V.B. (\email{vincent.bourrier@unige.ch})}

\institute{
Observatoire de l'Universit\'e de Gen\`eve, 51 chemin des Maillettes, 1290 Sauverny, Switzerland
\and 
Physikalisches Institut, Sidlerstr. 5, University of Bern, 3012, Bern, Switzerland
}
   
   \date{} 
 
  \abstract
{Kepler-444 provides a unique opportunity to probe the atmospheric composition and evolution of a compact system of exoplanets smaller than the Earth. Five planets transit this bright K star at close orbital distances, but they are too small for their putative lower atmosphere to be probed at optical/infrared wavelengths. We used the Space Telescope Imaging Spectrograph instrument onboard the Hubble Space Telescope to search for the signature of the planet's upper atmospheres at six independent epochs in the Lyman-$\alpha$ line. We detect significant flux variations during the transits of both Kepler-444\,e and \,f ($\sim$20\%), and also at a time when none of the known planets was transiting ($\sim$40\%). Variability in the transition region and corona of the host star might be the source of these variations. Yet, their amplitude over short time scales ($\sim$2--3\,hours) is surprisingly strong for this old (11.2$\pm$1.0\,Gyr) and apparently quiet main-sequence star. Alternatively, we show that the in-transits variations could be explained by absorption from neutral hydrogen exospheres trailing the two outer planets (Kepler-444\,e and \,f). They would have to contain substantial amounts of water to replenish such hydrogen exospheres, which would reveal them as the first confirmed ocean-planets. The out-of-transit variations, however, would require the presence of a yet-undetected Kepler-444\,g at larger orbital distance, casting doubt on the planetary origin scenario. Using HARPS-N observations in the sodium doublet, we derived the properties of two Interstellar Medium clouds along the line-of-sight toward Kepler-444. This allowed us to reconstruct the stellar Lyman-$\alpha$ line profile and to estimate the XUV irradiation from the star, which would still allow for a moderate mass loss from the outer planets after 11.2\,Gyr. Follow-up of the system at XUV wavelengths will be required to assess this tantalizing possibility.}


   \maketitle

\section{Introduction}
\label{intro} 

\subsection{Characterizing small planets through evaporation}

About a quarter of the known exoplanets orbit at short distances ($\approxinf$0.1\,au) from their star (from the Exoplanet Encyclopaedia in December 2016; \citealt{Schneider2011}). Heating by the stellar energy can lead to the expansion of their upper atmospheric layers and their eventual escape. Because of its expansion, the upper atmosphere produces a deeper absorption than the planetary disk alone when observed in the UV, in particular in the stellar Lyman-$\alpha$ (Ly-$\alpha$) line of neutral hydrogen (e.g., \citealt{VM2003}, \citealt{Lecav2012}). Super Earths and smaller planets display a large diversity in nature and composition (eg, \citealt{Seager2007}, \citealt{Rogers2010}, \citealt{Fortney2013}), which can only be investigated through observations of their atmosphere. The reduced scale height of the lower atmospheric layers makes them difficult to probe in the visible and the infrared. In contrast, very deep UV transit signatures can be produced by the upper atmospheres of small planets. The warm Neptune GJ436b, which is the lowest-mass planet found evaporating to date (\citealt{Ehrenreich2015}), shows transit absorption depths up to 60\% in the Ly-$\alpha$ line. The formation of such an extended exosphere is due in great part to the low mass of GJ436b and the gentle irradiation from its M-dwarf host (\citealt{Bourrier2015_GJ436}, \citealt{Bourrier2016}). Few attempts have been made to detect atmospheric escape from Earth-sized planets, and Ly-$\alpha$ transit observations of the super Earth 55 Cnc e (\citealt{Ehrenreich2012}) and HD\,97658 b (\citealt{Bourrier2016_HD976}) showed no evidence for hydrogen exospheres. In the case of 55 Cnc e, this non-detection hinted at the presence of a high-weight atmosphere -- or the absence of an atmosphere -- recently supported by the study of its brightness map in the IR (\citealt{Demory2016}). Understanding the conditions that can lead to the evaporation of Earth-size planets will be necessary to determine the stability of their atmosphere, and their possible habitability. For example, large amounts of hydrogen in the upper atmosphere of a close-in terrestrial planet could indicate the presence of a steam envelope being photo-dissociated, and replenished by evaporating water oceans (\citealt{Jura2004}, \citealt{leger2004}). \\

\subsection{The Kepler-444 system} 

Five planets smaller than the Earth were detected at close orbital distances (P$<$10\,days) around Kepler-444\,A, in the oldest known planetary system (\citealt{Campante2015}; 11.2$\pm$1.0\,Gy). Given the shallowness of the planetary transits, their atmospheres are undetectable with current facilities in the optical/nIR. The irradiation from their K0-type host star could however lead to the extension of their upper atmosphere, making them observable at UV wavelengths. A bright and nearby star (V=8.9; d=35.7\,pc; \citealt{VanLeeuwen2007}), Kepler-444\,A has also a high systemic velocity that favors observations in the Ly-$\alpha$ line. It is part of a triple-star system, with a bound M-dwarf pair on a highly eccentric, 430\,yr-long orbit.

We present in this paper a comprehensive study of the Kepler-444 system based on multi-epochs observations, which are presented in Sect.~\ref{sec:data_red}. Ly-$\alpha$ observations with the Hubble Space Telescope (HST) are analyzed in Sect.~\ref{sec:ana_Lalpha}. In Sect.~\ref{sec:ISM_carac} we measure the properties of the interstellar medium (ISM) in the direction of Kepler-444\,A, which are then used in Sect.~\ref{sec:ISM_XUV} to estimate the intrinsic Ly-$\alpha$ line and XEUV spectrum of the star. Those results are used in Sect.~\ref{sec:evap_planets} to study atmospheric escape from the Kepler-444 planets, comparing the Ly-$\alpha$ spectra with numerical simulations obtained with the EVaporating Exoplanet (EVE) code. We present our conclusions in Sect.~\ref{sec:conclu}. All system properties required for our analysis were taken from \citet{Campante2015} and \citet{Dupuy2016}.\\

\section{Observations and data reduction}
\label{sec:data_red} 

\subsection{Ly-$\alpha$ observations} 

We observed Kepler-444\,A in the H\,{\sc i} Ly-$\alpha$ line (1215.6702\,\AA) with the Space Telescope Imaging Spectrograph (STIS) instrument on board the HST. Six visits were obtained at different epochs (Table.~\ref{tab:log}) in the frame of GO Program 14143 (PI: V.~Bourrier). The configuration of the planetary system in each visit is shown in Fig.~\ref{fig:orb_cov}. Our main objective was to search for signatures of planetary hydrogen escape through transit spectroscopy. Visits A, B, C, and F were thus scheduled during the transits of one or several planets. Visit D and E were scheduled out of any planet transit, to measure the reference Ly-$\alpha$ line. We note that Visit D was initially scheduled as a five-orbit visit. However, a Fine Guidance Sensors failure of the HST prevented the reacquisition of the guide stars after the first orbit. No data was subsequently obtained, either because the target star drifted outside of the slit, or because the shutter did not reopen. Five new HST orbits were granted by the HST Telescope Time Review Board, and rescheduled as Visits E and F. 

\begin{figure}     
\includegraphics[trim=0cm 0.cm 0cm 0cm,clip=true,width=\columnwidth]{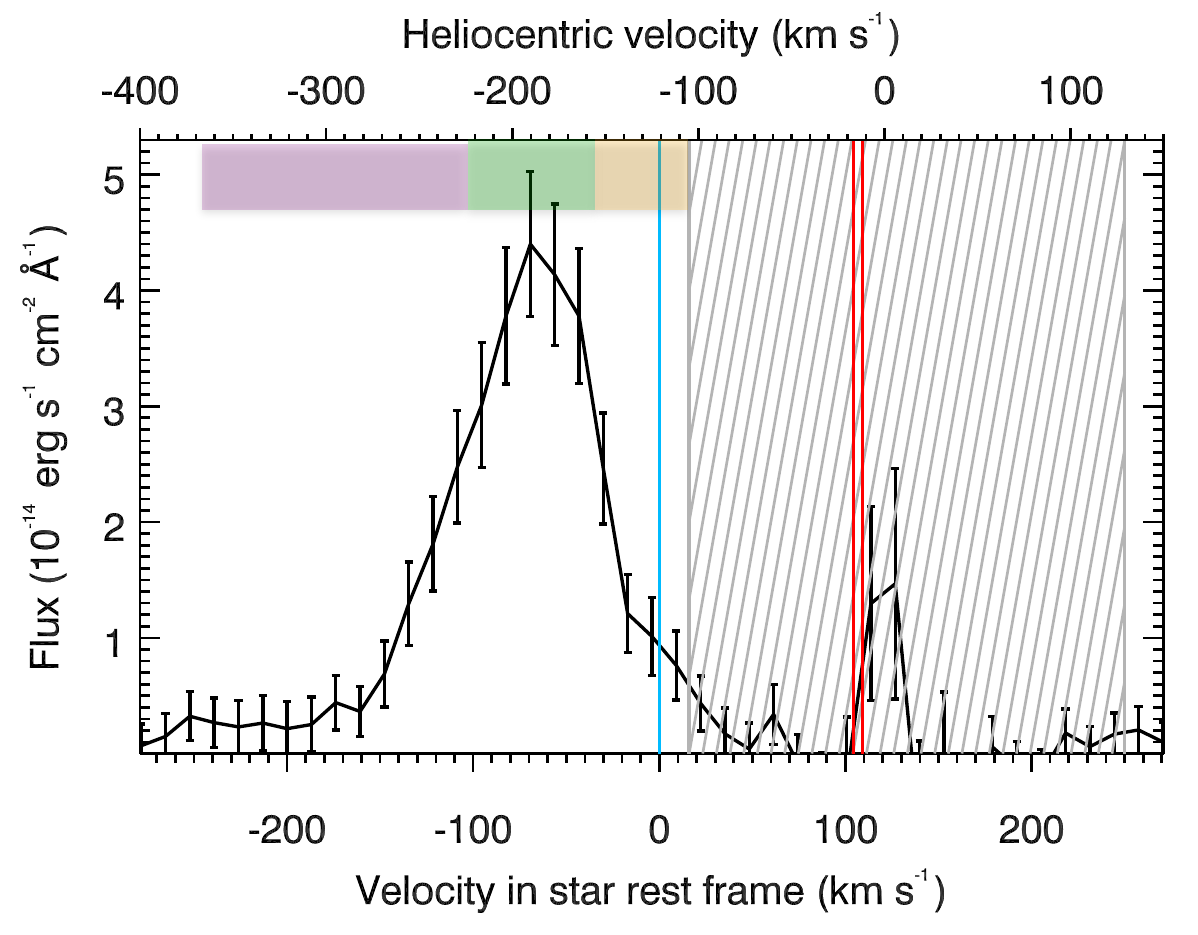}
\caption[]{Ly-$\alpha$ line spectrum of Kepler-444A, plotted as a function of Doppler velocity in the star rest frame (lower axis) and in the heliocentric rest frame (upper axis). The black line shows a typical measurement of the stellar line with HST/STIS (first orbit of Visit E). The violet, green, and orange bands indicate the spectral ranges of the wing, peak, and core bands, respectively. The blue line indicates the velocity of the star, and the red lines the velocities of the two ISM components along the LOS. The red wing of the Ly-$\alpha$ line cannot be observed from Earth because of ISM absorption along the line of sight (LOS). The hatched region is further contaminated by geocoronal emission.}
\label{fig:helio_spectrum}
\end{figure}

Data obtained with the G140M grating were reduced into 1D spectra with the CALSTIS pipeline, which corrects for the geocoronal airglow emission superimposed with the stellar Ly-$\alpha$ emission (\citealt{VM2003}). Airglow contamination was limited by the use of STIS narrow slit of 52''$\times$0.05'', but the spectra were not properly corrected at heliocentric velocities near 0\,km\,s$^{-1}$ where the airglow is maximum and the stellar line is fully absorbed by the ISM (Fig.~\ref{fig:helio_spectrum}). Using different areas of the 2D images to build manually the airglow profile did not improve the correction. We thus excluded from our analysis the spectral ranges shown in Fig.~\ref{fig:grid_spec}, accounting for the variation in strength and position of the airglow with the epoch of observation. The excluded ranges overlap with the red wing of the stellar Ly-$\alpha$ line. The range of heliocentric radial velocities of clouds in the local ISM can be roughly approximated to $\pm$20\,km\,s$^{-1}$ (\citealt{Redfield_Linsky2008}) and their absorption is usually maximum in the core of the stellar line for nearby, low-velocity stars. While ISM absorption is quite strong for Kepler-444A (see Sect.~\ref{sec:Lalpha_rec}), the system is singular in its high-proper-motion toward the Sun. With a radial velocity of -121.4\,km\,s$^{-1}$ (\citealt{Dupuy2016}) the stellar Ly-$\alpha$ line is shifted blueward of the spectrum in the heliocentric frame. This allowed us to measure the blue wing and the core of the line, freed up from both ISM and airglow contamination (Fig.~\ref{fig:helio_spectrum}).

Analysis of the time-tagged data did not reveal any significant signature of the breathing effect known to affect STIS UV observations (e.g., \citealt{Bourrier2016_HD976}). The amplitude of the breathing variations can change between visits of the same target, and it may have been low at the time of our observations. The noise in the faint Ly-$\alpha$ line of Kepler-444\,A also likely dominates the breathing variations. In any case the orbit-to-orbit variations of the breathing are repeatable (e.g. \citealt{Brown2001}; \citealt{Sing2008a}; \citealt{Huitson2012}; \citealt{Ehrenreich2015}), and to mitigate any residual effect we used the spectra averaged over the full duration of each HST orbit. We note  that the bound M-dwarf pair Kepler-444\,BC is currently near its apastron, and lies $1\farcs8$ (66\,au) away from the primary star (\citealt{Dupuy2016}). Even if by mischance those faint M dwarfs had entered the $0\farcs05$-wide slit, their spectrum would have been located between about 60 to 70 pixels from the spectrum of Kepler-444\,A. No excess signal was measured in this region of the STIS 2D images, which is not in any case used to correct the stellar spectrum from the background.

\begin{table}[tbh]
\caption{Log of Kepler-444 Ly-$\alpha$ observations.}
\begin{tabular}{lccc}
\hline
\hline
\noalign{\smallskip}
Visit & Date & \multicolumn{2}{c}{Time (UT)}   \\
      &      & Start & End			          \\
\noalign{\smallskip}
\hline
A & 2015-11-18     & 16:08:00   &  22:59:26\\
B & 2015-12-07     & 10:05:12   &  15:19:27\\
C & 2015-12-17     & 03:53:55   &  09:09:19\\
D & 2016-05-05     & 03:52:30   &  04:22:22\\
E & 2016-06-06     & 08:43:33   &  13:53:12\\
F & 2016-07-08     & 21:06:08   &  21:36:00\\
\noalign{\smallskip}
\hline
\hline
\end{tabular}
\label{tab:log}
\end{table}

\subsection{HARPS-N observations}
\label{sec:data_Harps}

We obtained four spectra of Kepler-444A with HARPS-N in the visible range, to search for ISM absorption in the sodium doublet and assess stellar activity. Observations were obtained via a change request of the SPADES program (OPT16A49, PI: D. Ehrenreich). Each spectrum results from a 30-mn long exposure, with two spectra obtained on 2016, March 18, and two spectra on April, 11. We used Molecfit (an ESO tool, \citealt{Smette2015}, \citealt{Kausch2015}) to correct the spectra for telluric contamination. A telluric template was generated at the location and date of each observation, and adjusted to the observed spectra using the telluric lines of main visible absorbers (O$_{2}$ and H$_{2}$O; Allart et al. 2017, in prep). We limited the fit to specific regions, where we excluded small telluric lines, strong stellar lines, and blends of telluric and stellar lines. The entire observed spectra were then corrected using the derived telluric template and the Calctrans tool.\\

\section{Variations in the Ly-$\alpha$ line}
\label{sec:ana_Lalpha}

\begin{figure*}
\centering
\begin{minipage}[b]{\textwidth}   
\includegraphics[trim=0cm 7.75cm 0.95cm 15.35cm,clip=true,width=\textwidth]{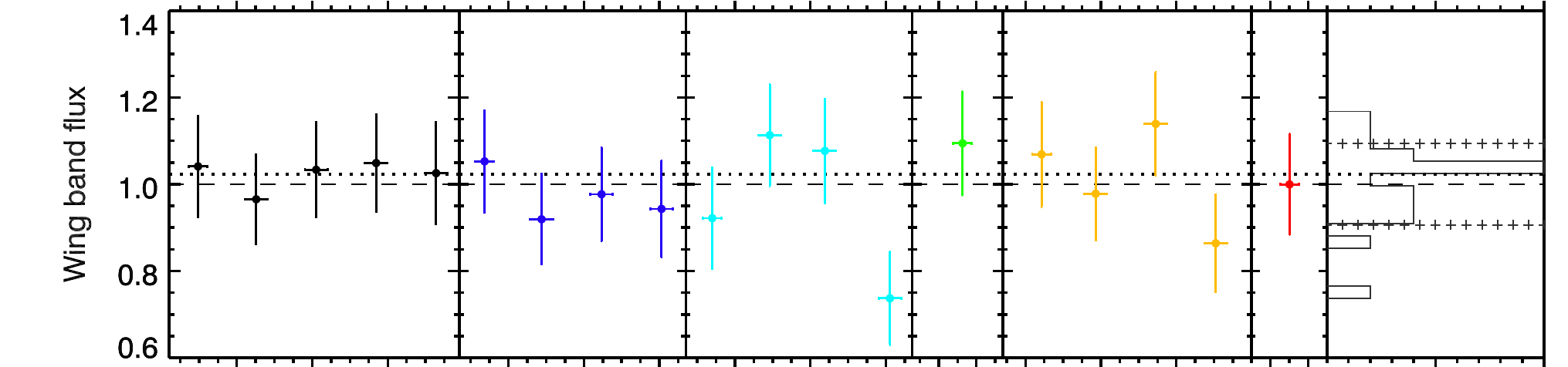}\\
\includegraphics[trim=0cm 7.75cm 0.95cm 15.45cm,clip=true,width=\textwidth]{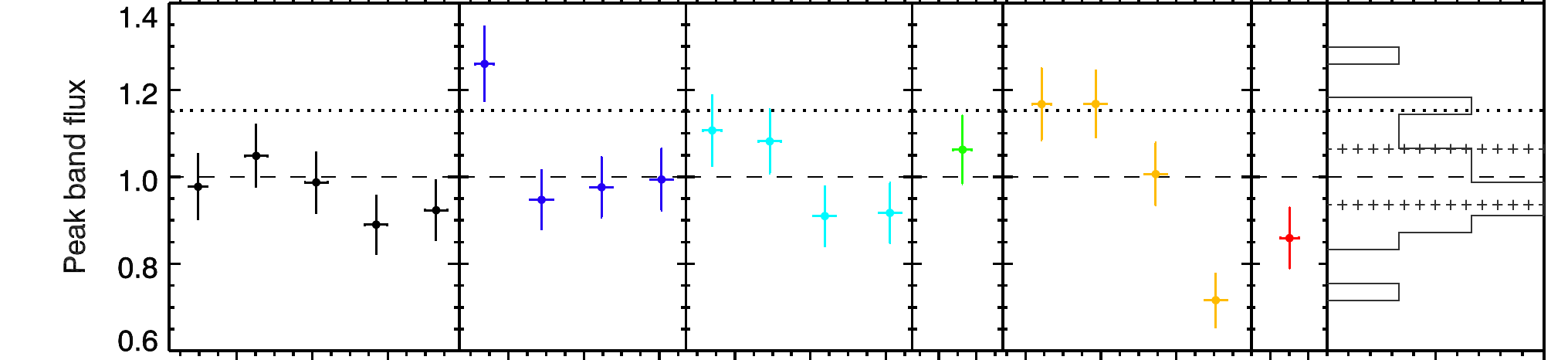}\\
\includegraphics[trim=0cm 6.1cm 0.95cm 15.45cm,clip=true,width=\textwidth]{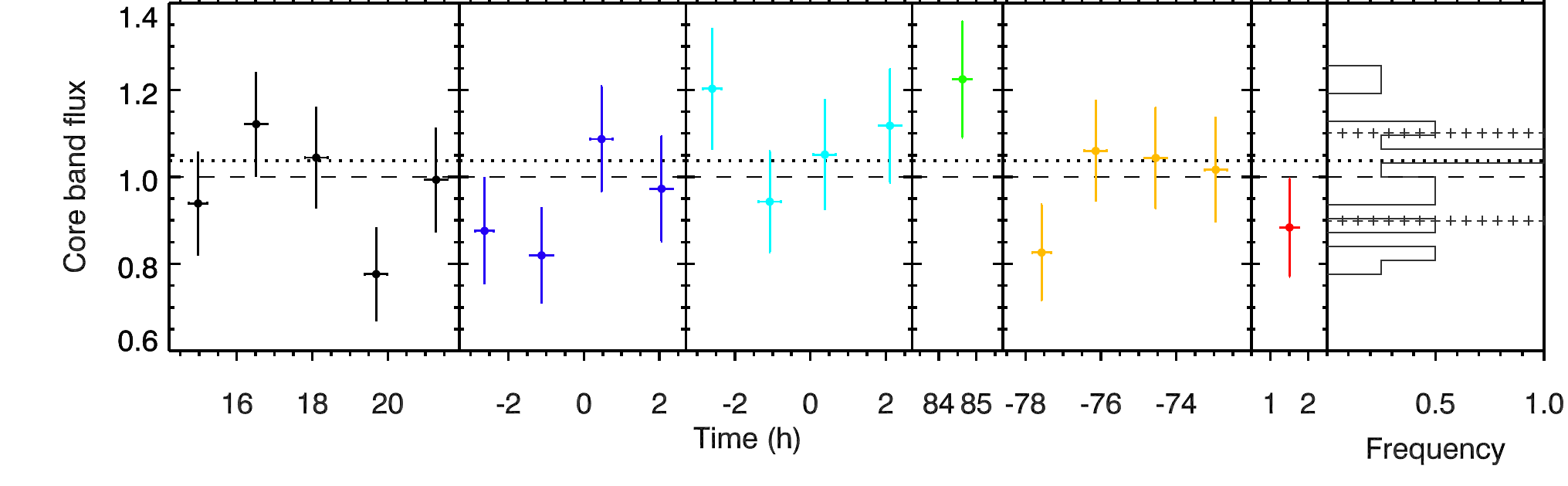}\\
\end{minipage}
\caption[]{Ly-$\alpha$ flux in Visits A (black), B (deep blue), C (light blue), D (green), E (orange), and F (red). Times are relative to the mid-transit of Kepler-444\,f. All spectra have been interpolated over a common wavelength table before being integrated in the complementary wing, peak, and core bands (see text), displayed from top to bottom. Fluxes have been normalized by their average over all visits (dashed line). The dotted line corresponds to the spectrum taken as reference for the out-of-transit line. Right panels show the flux distribution in each band. Cross-symbol lines delimit the 68\% confidence interval around the average flux, if all bands followed the same statistical variations as the wing band.}
\label{fig:light_curves}
\end{figure*}

In a first step, we compared all STIS spectra together to assess the variability in the Ly-$\alpha$ line. Significant flux variations were found in different spectral ranges and with varying amplitude depending on the orbit and the visit (Fig.~\ref{fig:grid_spec}). We found that we could identify three complementary bands with approximately the same behaviour, and their integrated flux over time is displayed in Fig.~\ref{fig:light_curves}. At high negative velocities in the ``wing'' band (-246 ; -102\,km\,s$^{-1}$ in the star rest frame) the line is very stable, with no significant flux variation within each visit and over the six epochs of observations (upper panel in Fig.~\ref{fig:light_curves}). Accordingly, the largest deviation between the flux measurements and their average is at 2.4$\sigma$, and the reduced $\chi^2$ between the flux measurements and their average is 0.8. Assuming that the measured spectra follow statistical variations dictated by Gaussian photon noise around an average spectrum, we used the rms of the flux in the wing band to estimate the dispersion expected in the ``peak'' (-102 ; -37\,km\,s$^{-1}$) and the ``core'' (-37 ; 16\,km\,s$^{-1}$) bands. We found that the actual dispersion in the peak band is larger than the statistical predictions (middle panel in Fig.~\ref{fig:light_curves}), with more than half of the measurements outside of the predicted 68\% interval. Those deviations from the average flux are significant (up to 30\% at the 4.6\,$\sigma$ level in Visit E) and the reduced $\chi^2$ between the flux measurements and their average reaches a high value of 3.2. The conclusion is less clear for the core band (reduced $\chi^2$ of 1.2, largest deviation at 2.1$\sigma$; see lower panel in Fig.~\ref{fig:light_curves}), although its variations might have a physical origin (Sect.~\ref{sec:pl_abs}). Hereafter we investigate whether these variations in the Ly-$\alpha$ line arise from stellar variability, or planetary absorption. \\

\subsection{Stellar variability}
\label{sec:st_var}

Different regions of the stellar atmosphere contribute to the global Ly-$\alpha$ line profile. The low-flux wings of the line are formed in the colder regions of the lower chromosphere. Regions of higher flux at the peaks of the line are emitted by the transition region between the upper chromosphere of the star and its corona. In this respect, the stability of Kepler-444\,A Ly-$\alpha$ line in the wing band would suggest a stable chromosphere, while the variations observed in the peak band could indicate a variable transition region and corona (see, e.g., the case of HD\,97658; \citealt{Bourrier2016_HD976}). However, it is puzzling that an 11.2\,Gy old, K-type star displays such strong variations over very short timescales. Indeed, the average stellar line remains quite stable between visits, over $\sim$8\,months of observations (Fig.~\ref{fig:light_curves}, Fig.~\ref{fig:grid_spec}). Yet the overall dispersion of the Ly-$\alpha$ flux in the peak band is about 10\%, higher than the expected statistical dispersion of $\sim$5\% (Sect.~\ref{sec:ana_Lalpha}), because of systematic orbit-to-orbit variations on the order of $\sim$20-30\% in Visits B, C, and E. The largest and most significant variation is observed in Visit E when the flux decreases by 40$\pm$6\% (6.5\,$\sigma$) from the beginning to the end of the visit, in only about two hours.\\
\citet{Llama2016} assessed the variability of the Sun at Ly-$\alpha$ over the timescales of exoplanet transits. They used disk-integrated observations of the Sun obtained with a 10\,s cadence over more than 3000 hours with the Solar Dynamics Observatory/Extreme Ultraviolet Variability Experiment instrument. Rebinning these observations to a cadence typical of HST/STIS science exposures, they found that the solar Ly-$\alpha$ variability does not exceed 3\%, with less than 5\% of the measurements showing more than 1.5\% variability. Kepler-444\,A thus displays a much larger variability with a higher frequency. A K dwarf like Kepler-444\,A can nonetheless be more active than the Sun, and we used the study of the transiting super-Earth HD\,97658 b made by \citet{Bourrier2016_HD976} to estimate the variability of its K1-type host star at Ly-$\alpha$. HD\,97658 could be as old as Kepler-444 (9.7$\pm$2.8\,Gyr; \citealt{Bonfanti2016}), and was observed at three epochs in the Ly-$\alpha$ line. The peaks of the line revealed systematic variations, attributed to an active upper chromosphere/corona, which reach a maximum of $\sim$15\% from one HST orbit to another. Although the small number of observations of HD\,97658 limits our assessment of its Ly-$\alpha$ variability, this further suggests that Kepler-444\,A show unusually large variations.\\
Finally, we measured the chromospheric activity indicators (S-index, \citealt{Vaughan1978}; H$\alpha$, \citealt{daSilva2011}) in the four HARPS-N spectra of Kepler-444\,A. The average log(R'$_{HK}$) of -5.179 (computed following \citealt{Astudillo2016}) indicates a relatively inactive star, and we found no significant change in activity over the month separating the two epochs of HARPS-N observations (Table~\ref{tab:activ}). This strengthen our conclusion that chromospheric activity is not the origin of the observed variations.\\

\begin{table}[tbh]
\caption{Evolution of Kepler-444\,A activity indicators.}
\label{tab:activ}
\begin{threeparttable}
\begin{tabular}{lcc}
\hline
\hline
\noalign{\smallskip}
Time (BJD)	& H$\alpha$	&	S-index \\
\noalign{\smallskip}
\hline
57490.71240 &	0.03596$\pm$3$\times$10$^{-5}$ & 0.13$\pm$0.10\\
57466.72836 &	0.03586$\pm$3$\times$10$^{-5}$ & 0.13$\pm$0.10\\
57490.73980 &	0.03602$\pm$3$\times$10$^{-5}$ & 0.14$\pm$0.11\\
57466.75003 &	0.03591$\pm$3$\times$10$^{-5}$ & 0.14$\pm$0.11\\
\noalign{\smallskip}
\hline
\hline
\end{tabular}
  \begin{tablenotes}[para,flushleft]
  Note: Uncertainties are derived from photon-noise alone.
  \end{tablenotes}
  \end{threeparttable}
\end{table}

\subsection{Planetary absorption}
\label{sec:pl_abs}

Neutral hydrogen exospheres have been detected around all\footnote{Except for WASP-12b (\citealt{Fossati2010}), which is too far away for Ly-$\alpha$ transmission spectroscopy.} known evaporating planets (HD\,209458b, \citealt{VM2003}; HD\,189733b, \citealt{Lecav2010}, \citealt{Lecav2012}; 55\,Cnc b, \citealt{Ehrenreich2012}; GJ\,436b, \citealt{Ehrenreich2015}). The interplay of stellar gravity, radiation pressure and stellar wind interactions shapes the escaping gas into comet-like tails, which absorb in the blue wing of the Ly-$\alpha$ line because of their acceleration away from the star, and transit long after the planetary disk because of their spatial extension (e.g., \citealt{Bourrier_lecav2013}, \citealt{Kislyakova2014}, \citealt{Bourrier2016}). We hypothesize that the variations observed in the Ly-$\alpha$ line of Kepler-444\,A could be explained by the transit of several exospheres absorbing in the peak, and possibly the core, bands. The stability of the flux in the wing band would be explained by the fact that hydrogen atoms cannot be accelerated beyond a maximum velocity by mechanisms such as radiation pressure or charge-exchange (\citealt{Holmstrom2008}, \citealt{Ekenback2010}, \citealt{Bourrier_lecav2013}).

\begin{figure}
\centering
\includegraphics[trim=0cm 5.5cm 3cm 10.5cm,clip=true,width=\columnwidth]{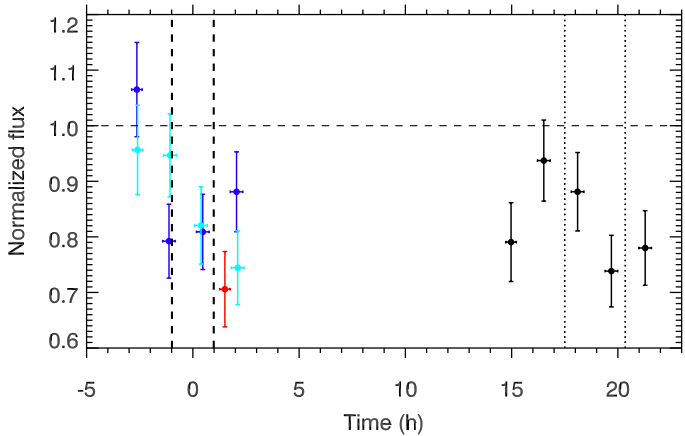}\\
\caption[]{Ly-$\alpha$ flux integrated in the range [-89;-37]\,km\,s$^{-1}$, phase-folded using planet 'f' ephemeris. Dashed vertical lines indicate the contacts of planet 'f', and dotted vertical lines the contacts of planet 'e' in Visit A. Color code is the same as in Fig.~\ref{fig:light_curves}.}
\label{fig:light_curve_plf}
\end{figure}

The first orbits in Visits B and C, as well as the single orbit in Visit D and the two first orbits in Visit E, show similar flux levels (Fig.~\ref{fig:light_curves}, Fig.~\ref{fig:grid_spec}). None of the five planets in the system were transiting the star during these orbits (Fig.~\ref{fig:orb_cov}), and we thus averaged their spectra to create a reference for the unocculted stellar line. We then compared the other spectra to this reference, and found that the flux decrease observed in the peak band in Visits B, C and F could be correlated with the transit of a hydrogen exosphere trailing planet 'f'. To illustrate this, we rephased the observations using planet 'f' ephemeris in Visit A (Fig.~\ref{fig:light_curve_plf}). The resulting light curve suggests that the putative exosphere of planet 'f' could still be transiting the star about 15\,h after the optical transit, just before the transit of planet 'e'. The flux decrease observed during planet 'e' transit would then be explained by the combined absorptions of planets 'f' and planet 'e' exospheres. We study this scenario in Sect.~\ref{sec:sim_EVE}. 

With orbital periods of 7.7 and 9.7\,days, Kepler-444\,e and Kepler-444\,f might receive enough energy to yield a substantial hydrogen escape, while being far enough from the star for the resulting exosphere to expand and survive photoionization. This will be investigated in Sect.~\ref{sec:esc_rate}. Closer-in planets are more likely to have lost all of their hydrogen, although the flux variations observed in the core band might arise from additional absorption by planets 'b' and 'c'. We also note that the third and fourth spectra in Visit E display a significant flux decrease in the peak band, even though all known planets in the system are far before their respective transits. We address this surprising variation in Sect.~\ref{sec:444g}.

\section{Characterization of the ISM toward Kepler-444}
\label{sec:ISM_carac}

Measurements of the Ly-$\alpha$ line at Earth can usually be used to estimate the ISM properties along the LOS, a necessary step to reconstruct the intrinsic stellar Ly-$\alpha$ line (e.g., \citealt{Bourrier2013}, \citealt{youngblood2016}). This is not possible in the case of Kepler-444\,A, however, because the red wing of its Ly-$\alpha$ line is fully absorbed by the ISM and further contaminated by the airglow (Fig.~\ref{fig:helio_spectrum}). \\

\subsection{Interstellar sodium absorption}
\label{sec:ISM_sodium}

\begin{figure}
\centering
\includegraphics[trim=0.5cm 1.5cm 1.2cm 4.5cm,clip=true,width=\columnwidth]{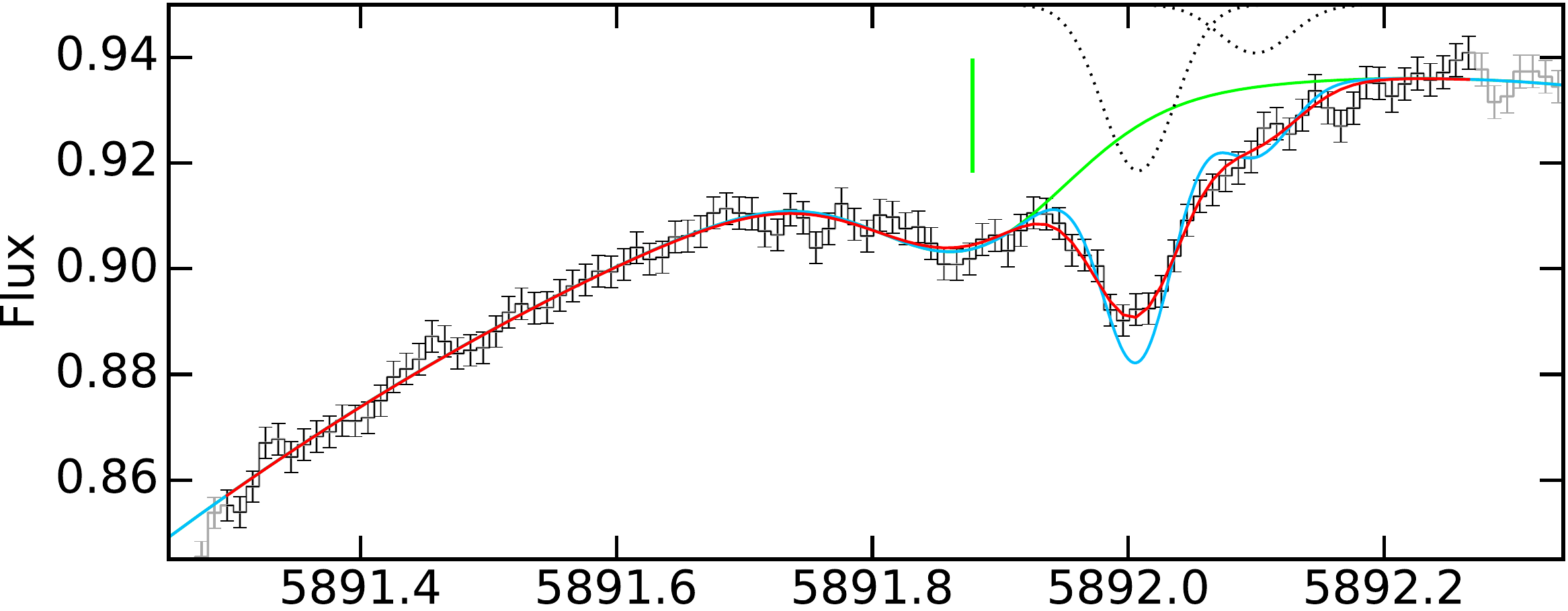}\\
\includegraphics[trim=0.5cm 0.5cm 1.2cm 4.5cm,clip=true,width=\columnwidth]{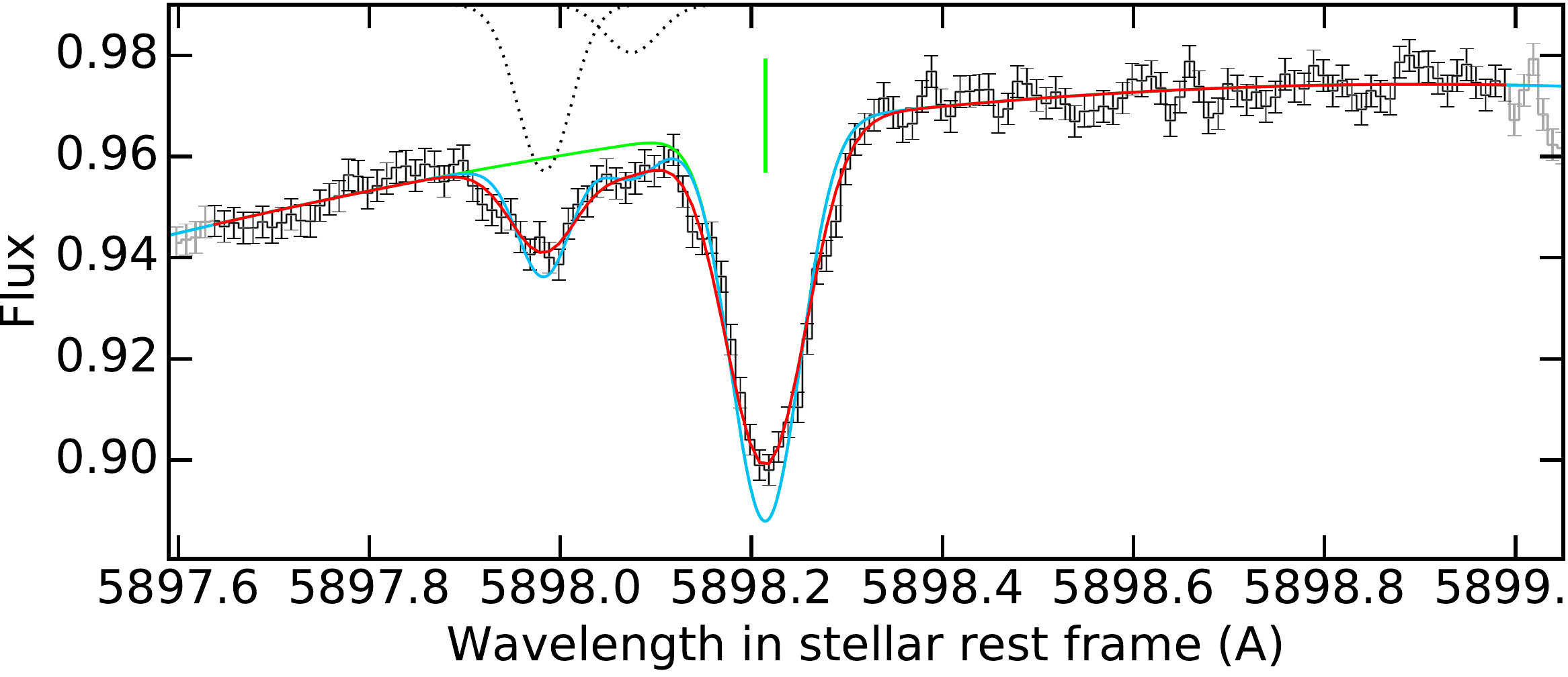}\\
\caption[]{Spectrum of Kepler-444\,A in the regions of interstellar absorption from the Na\,{\sc i}\,D2 (top) and Na\,{\sc i}\,D1 (bottom) lines. Our best-fit for the intrinsic stellar spectrum (green line) yields the blue spectrum after absorption by two ISM components (absorption profiles displayed as dotted black lines, with the deeper A component blueshifted with respect to the shallower B component). Its convolution with HARPS-N instrumental profile yields the red curve, that was compared with the data over the black points. Note that the Na\,{\sc i}\,D2 absorption line from the A component, and the Na\,{\sc i}\,D1 absorption line from the B component, are blended with stellar Fe\,{\sc i} lines, whose centers are indicated by vertical green lines.}
\label{fig:ISM_plots_Na}
\end{figure}

We obtained an independent measurement of the ISM properties through the analysis of interstellar absorption from the Na\,{\sc i} doublet (Na\,{\sc i}\,D2 at 5889.95\,\AA\, and Na\,{\sc i}\,D1 at 5895.92\,\AA) in the HARPS-N spectra. We did not detect any clear ISM absorption in the Ca\,{\sc ii} h (3968.47\,\AA) \& k (3933.66\,\AA), mainly because of the much lower stellar flux and signal-to-noise ratio (SNR) in this region of the spectra. The four HARPS-N spectra reduced in Sect.~\ref{sec:data_Harps} were coadded before being adjusted with a theoretical model. Fourth-order polynomials were used to represent the stellar continuum in the range of each sodium line. Because of the high radial velocity of Kepler-444, ISM sodium absorbs in the red wings of the stellar sodium lines (Fig.~\ref{fig:ISM_plots_Na}). We found that the ISM sodium lines were blended with stellar Fe\,{\sc i} lines at 5891.88\,\AA\, and 5898.21\,\AA, which were included in the model. Voigt profiles were used for stellar and interstellar lines. The free parameters of the ISM are the column density log$_{10}$\,N(Na\,{\sc i}), the Doppler broadening parameter b\,(Na\,{\sc i}), and the heliocentric radial velocity $\gamma$(Na\,{\sc i})$_{/\mathrm{Sun}}$. The final model was obtained after convolution with HARPS-N line spread function (LSF) and was adjusted simultaneously to both sodium lines using the Bayesian Information Criterion (BIC; \citealt{Liddle2007}) as merit function. We obtained a significantly better fit when including a second ISM component (BIC difference of 18). Because the Doppler broadening of this second component was not well constrained, we fitted the same parameter to both ISM components (labelled A and B). Our best fit is displayed in Fig.~\ref{fig:ISM_plots_Na}, with the corresponding ISM properties in Table~\ref{tab:results_ISM_La}.\\

\subsection{Properties of the ISM toward Kepler-444}
\label{sec:ISM_K444}

The LISM Kinematic Calculator\footnote{\mbox{\url{http://sredfield.web.wesleyan.edu/}}} predicts that the LOS toward Kepler-444\,A crosses the Local Interstellar Cloud (LIC) and Mic cloud. The velocities we derived for the A and B components are consistent with those predicted for the Mic (-18.0$\pm$1.4\,km\,s$^{-1}$) and the LIC (-10.3$\pm$1.4\,km\,s$^{-1}$), respectively. However, our broadening parameter is noticeably lower than the ranges expected for the Mic and the LIC (3.2--5.0 and 2.3--3.5\,km\,s$^{-1}$, respectively, from the clouds temperature and turbulent velocity in \citealt{Redfield_Linsky2008}). Furthermore, we found that the LOS toward Kepler-444 actually stands near the edges of the LIC and Mic cloud boundaries estimated by \citet{Redfield_Linsky2008}). The uncertainty in these clouds morphologies, probed with a small number of LOS in the sky region toward Kepler-444, suggests that its LOS may probe colder clouds extending beyond the LIC and the Mic clouds. \\
\citet{Welty1996} identified three ISM cloud components (Clouds 1, 2, and 3) in the direction of $\delta$ Cyg (galactic coordinates $l$=78.7$^{\circ}$, $b$=10.2$^{\circ}$), which stands very close to the LOS toward Kepler-444 ($l$=73.4$^{\circ}$, $b$=12.9$^{\circ}$). Using interstellar Ca\,{\sc ii} absorption, they derived heliocentric velocities of about -18.8, -16.3, and -9.6\,km\,s$^{-1}$, respectively (given the broadness and shallowness of the Ca\,{\sc ii} absorption from Clouds 2 and 3, it is likely that the uncertainty on their velocity is larger than the standard value of 0.3\,km\,s$^{-1}$ given by \citealt{Welty1996}). Using their measurement of Ca\,{\sc ii} Doppler broadening for Clouds 1, 2, and 3, and assuming either pure thermal broadening or turbulent broadening, we estimate Na\,{\sc i} Doppler broadening in the ranges 0.5--0.6, 3--4, and 2.6--3.4\,km\,s$^{-1}$, respectively. The velocity and broadening of the Mic cloud and the LIC are thus consistent with the properties of Clouds 2 and 3, respectively, but not with the colder Cloud 1. While the properties of this cloud are not consistent with those derived for our A and B components either, it strengthens the idea that the LOS toward Kepler-444 could cross colder clouds than the Mic and the LIC. \\
In fact, \citet{Johnson2015} studied the ISM in the \textit{Kepler} search volume, measuring Na\,{\sc i} and K\,{\sc i} absorption toward 17 targets, and predicted that Kepler-444\,A could lie within two clouds (Clouds II and III) denser and farther away from the Sun than the LIC. Our value for the ISM Doppler broadening is within the similar ranges derived by \citet{Johnson2015} for these clouds (b\,(Na\,{\sc i})$\sim$[0.2--2.9]\,km\,s$^{-1}$). The velocity of our A component is within the range they obtained for Cloud II ([-17.2 ; -14]\,km\,s$^{-1}$), and the velocity of our B component is slightly lower than their range for Cloud III ([-11.4 ; -10.8]\,km\,s$^{-1}$). Finally, we derived a range of sodium column densities between the Sun and Kepler-444\,A, using \citet{Johnson2015} measurements for Clouds II and III. After excluding one outlying value among each sample used to probe those clouds, we found that our column density for the A component falls in the middle of the range estimated for Cloud II (log$_{10}$\,N(Na\,{\sc i})$\sim$9.8--11.5), while our column density for the B component is at the lower end of the range estimated for Cloud III (log$_{10}$\,N(Na\,{\sc i})$\sim$9.8--11.0). The difference in velocity and column density between our B component and cloud III might be caused by the absence of sample stars for this cloud at distances similar to that of Kepler-444. Nonetheless, there is a very good agreement between our measurements and those from \citet{Johnson2015}, and we thus conclude to the detection of two ISM components toward Kepler-444 consistent with their Clouds II and III.\\

\begin{table}
\centering
\caption{Properties of Kepler-444\,A XUV spectrum and ISM absorption.}
\label{tab:results_ISM_La}
\begin{threeparttable}
\begin{tabular}{lccc}
\hline
\hline
\noalign{\smallskip}    
\textbf{Parameter} & \multicolumn{2}{c}{\textbf{ISM components}} & \textbf{Units}\\ 
\noalign{\smallskip}
                   &  A &  B  &                     \\  
\hline  
\noalign{\smallskip}
$\gamma_{/Sun}$  & -16.8$\pm$0.1  & -12.1$\pm$0.4 & km\,s$^{-1}$   \\
$b$(Na\,{\sc i})   &  \multicolumn{2}{c}{2.00$\pm$0.17} & km\,s$^{-1}$\\                   
log$_{10}$\,N(Na\,{\sc i}) &  10.24$\pm$0.02   & 9.70$\pm$0.06  & cm$^{-2}$   \\                       
log$_{10}$\,N(H\,{\sc i}) &  18.54$\pm$0.24  & 18.00$\pm$0.24 & cm$^{-2}$   \\   
\noalign{\smallskip} 
\hline
\hline
\noalign{\smallskip} 
\textbf{Parameter} & \multicolumn{2}{c}{\textbf{XUV stellar spectrum}}  \\                       
\noalign{\smallskip}
\hline  
\noalign{\smallskip}
F$_{\mathrm{Ly}\alpha}$ &  \multicolumn{2}{c}{3.1$\stackrel{+1.4}{_{-0.6}}$}  \\
F$_{\mathrm{X}}$\,(5-100\,\AA) & \multicolumn{2}{c}{$\sim$0.1}  \\ 
F$_{\mathrm{EUV}}$\,(100-912\,\AA) & \multicolumn{2}{c}{1.3$\stackrel{+2.6}{_{-1.1}}$} \\ 
$\Gamma_{\mathrm{ion}}$ & \multicolumn{2}{c}{0.7$\stackrel{+1.1}{_{-0.4}}$}  \\
\noalign{\smallskip}
\hline
\hline
\end{tabular}
  \begin{tablenotes}[para,flushleft]
  Note: ISM and Ly-$\alpha$ line properties are derived from the fits to HARPS-N and HST/STIS spectra. The hydrogen column density for the B component is set by the ratio between the B and A components for sodium. Fluxes (in erg\,cm$^{-2}$\,s$^{-1}$) and photoionization rate (in 10$^{-7}$\,s$^{-1}$) are given at 1\,au from the star.
  \end{tablenotes}
  \end{threeparttable}
\end{table}       


\section{Intrinsic Ly-$\alpha$ line and EUV emission}
\label{sec:ISM_XUV}

We used the ISM properties measured in Sect.~\ref{sec:ISM_sodium} to reconstruct the intrinsic stellar Ly-$\alpha$ line. In turn, this reconstructed profile can be used to estimate the stellar EUV emission (Sect.~\ref{sec:xuv}) and to calculate the effects of radiation pressure on hydrogen exospheres (Sect.~\ref{sec:sim_EVE}). \\

\subsection{Ly-$\alpha$ line}
\label{sec:Lalpha_rec}

\begin{figure}     
\includegraphics[trim=0.cm 5.5cm 0cm 10.9cm,clip=true,width=\columnwidth]{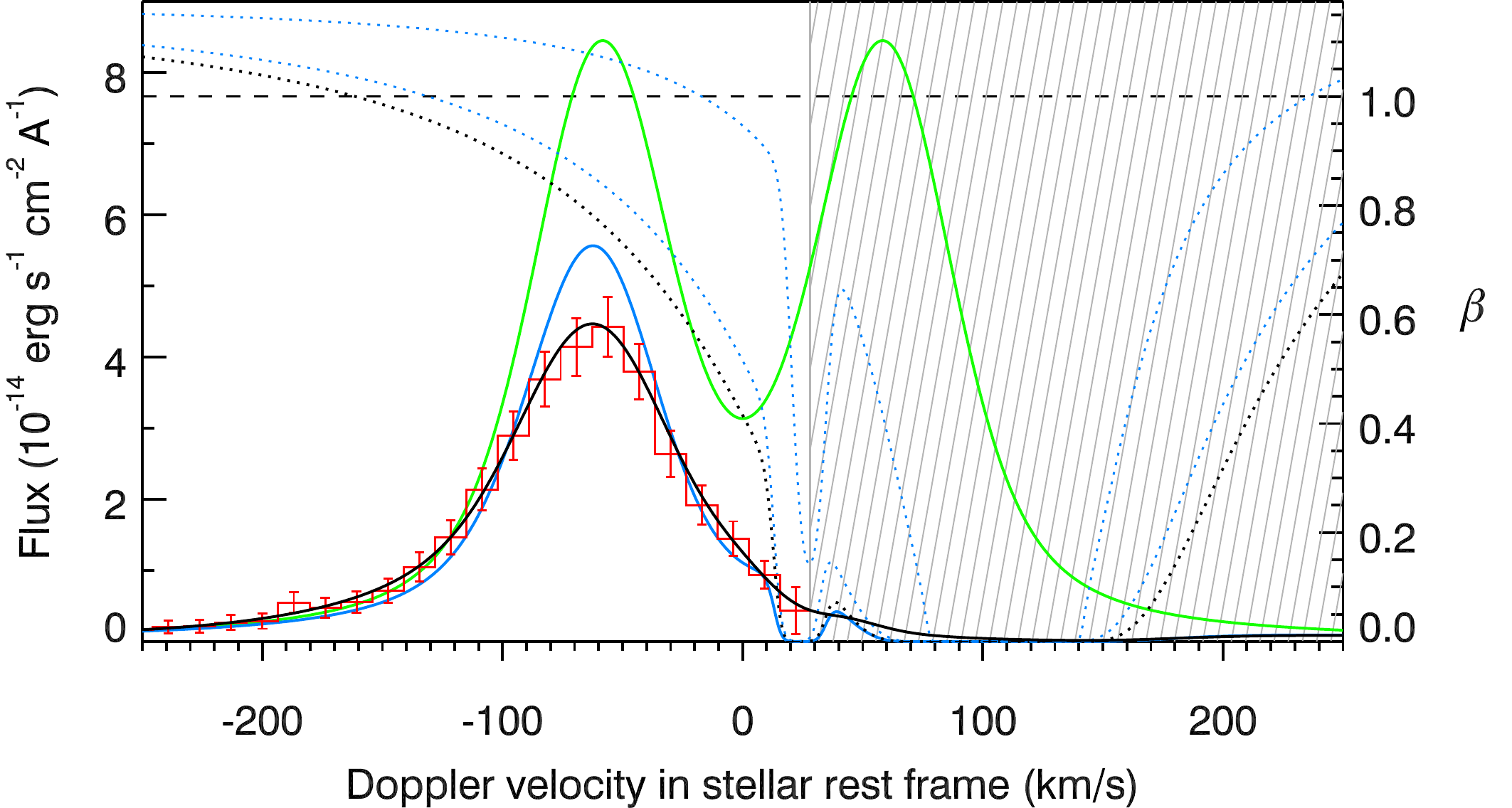}
\caption[]{Ly-$\alpha$ line profile of Kepler-444A. The green line shows the theoretical intrinsic stellar emission line profile scaled to the Earth distance. The absorption profiles of the two ISM components toward Kepler-444\,A are plotted as dashed blue lines, with their cumulated profile shown as a dashed black line. The solid blue line shows the Ly-$\alpha$ line profile after absorption by the ISM. The solid black line shows it convolution with STIS LSF, before it is compared to the out-of-transit spectrum shown as a red histogram. Hatched region was excluded from the fit. Right axis shows the ratio $\beta$ between radiation pressure and stellar gravity.}
\label{fig:theo_spectrum}
\end{figure}

We reconstructed the intrinsic Ly-$\alpha$ profile using the method detailed, e.g., in \citet{Bourrier2015_GJ436}. A model profile of the stellar line is absorbed by the ISM and convolved by STIS LSF. A Metropolis-Hasting Markov chain Monte Carlo algorithm is used to find the best fit between the model spectrum and the out-of-transit spectrum, which are compared over the range shown in Fig.~\ref{fig:theo_spectrum}. The heliocentric velocity of the star was fixed to -121.4\,km\,s$^{-1}$. ISM opacity is modeled as the combination of two Voigt profiles for the atomic hydrogen and deuterium, with a fixed D\,{\sc i}/H\,{\sc i} ratio of 1.5$\times$10$^{-5}$ (e.g., \citealt{Hebrard_Moos2003}, \citealt{Linsky2006}). The hydrogen column density of the A component was used as a free parameter, and we fixed the column density ratio between the B and A components to the value derived from the fit to the interstellar sodium (Table.~\ref{tab:results_ISM_La}). The results of this fit were also used to fix the radial velocities and Doppler broadening of the two ISM components. We note that the ISM Doppler broadening includes a thermal and non-thermal (turbulent) contribution (e.g., \citealt{Redfield2004}). It is not possible to distinguish between those two contributions from the absorption of a single element (here, sodium). We thus assumed there is no turbulent broadening, in which case the sodium line broadening yields a thermal broadening of $\sim$9.5\,km\,s$^{-1}$ for hydrogen. We checked that the opposite assumption (no thermal broadening) had no significant influence on the fit. This is consistent with the large hydrogen column density that results in the observed spectrum being mainly affected by the damping wings of the ISM absorption (Fig.~\ref{fig:theo_spectrum}). We tried different combinations of Gaussian and Voigt profiles for the intrinsic stellar Ly-$\alpha$ line, using the BIC as merit function. We found that the observations could be well reproduced using either a single-peaked (SP) Voigt profile or a double-peaked (DP) Voigt profile, with no strong statistical evidence for one model against the other ($\Delta$BIC=2.3 in favor of the DP profile). Nonetheless, there are several reasons in favour of the second scenario, which we detail below.\\

The peak and integrated fluxes of the intrinsic Ly-$\alpha$ line are significantly larger for the SP profile, which requires a stronger ISM absorption to explain the observed Ly-$\alpha$ spectrum. The ISM hydrogen column density associated to the SP profile (log$_{10}$\,$N$(H\,{\sc i}) = 19.21$\pm$0.07 for the A component) is larger by about one order of magnitude than for the DP profile (18.54$\pm$0.24), and larger than the range of values expected for a star at a distance of $\sim$35.7\,pc (Fig. 14 in \citealt{Wood2005}). We used the sodium column density derived in Sect.~\ref{sec:ISM_sodium} for the A component (log$_{10}$\,N(Na\,{\sc i}) =  10.24$\pm$0.02) to provide an independent estimate of the hydrogen column density. Using \citet{Ferlet1985} formula we derived log$_{10}$\,$N$(H\,{\sc i}) = 18.6, in very good agreement with the value obtained for the DP Ly-$\alpha$ line. \citet{Welty_Hobbs2001} suggested that \citet{Ferlet1985} relation could underestimate the hydrogen column density when log$_{10}$\,N(Na\,{\sc i}) $\approxinf$ 11, but we found that the relation derived by \citet{Welty_Hobbs2001} is only based on upper limits on the hydrogen column density for sodium column density below $\sim$10.3.\\
The combined effects of rotational line broadening and macroturbulence take a value of $\sim$2.2\,km\,s$^{-1}$ for Kepler-444\,A (\citealt{Campante2015}). Assuming a low macroturbulence of 1\,km\,s$^{-1}$ (macroturbulence velocity at T$_\mathrm{eff}$ = 5046\,K ranges between about 1 and 4\,km\,s$^{-1}$, \citealt{Ryabchikova2016}), we derived an upper limit on the rotational period of 32\,days. Even with a moderate stellar inclination, it is thus likely that Kepler-444\,A has a rotational period larger than 25 days, and we used the relation between the integrated Ly-$\alpha$ flux and stellar temperature derived for this class of rotators by \citet{Linsky2013}. It yields F$_{\mathrm{Ly}\alpha}\sim$6.2\,erg\,cm$^{-2}$\,s$^{-1}$ at 1\,au from the star (with a dispersion of about 90\%), significantly lower than the flux derived for the SP profile (22$\stackrel{+11}{_{-7}}$\,erg\,cm$^{-2}$\,s$^{-1}$) but consistent with the flux derived for the DP profile (3.1$\stackrel{+1.4}{_{-0.6}}$\,erg\,cm$^{-2}$\,s$^{-1}$).\\
Finally, a DP Ly-$\alpha$ line profile results from self-absorption in the upper stellar chromosphere, and is thus linked to the temperature-pressure profile of the chromosphere and the spectral type of the star. We compared the shape of the Ly-$\alpha$ line profiles reconstructed for G, K, and M-type stars by \citet{Wood2005} (and references therein), \citet{Ehrenreich2012}, \citet{Bourrier_lecav2013}, \citet{Bourrier2015_GJ436}, and \citet{Bourrier2016_HD976}. The Ly-$\alpha$ line profiles reconstructed for all M dwarfs (six) display a SP profile. This absence of self-reversal is further supported by the observation of the red dwarf Kapteyn's star (\citealt{Guinan2016}), whose Ly-$\alpha$ line - free of ISM absorption thanks to the high redshift of the star - displays a SP profile. Over 19 K dwarfs (nearly all of them with spectral types K0, K1, or K2), we found five Ly-$\alpha$ line profiles with no self-reversals, five with deep self-reversals, and the others with self-reversals of various depths. The Ly-$\alpha$ line profiles reconstructed for 24 G-type stars all display a DP profile like the Sun (e.g., \citealt{Lemaire2002}). This comparison highlights that K dwarfs have intermediate chromospheric structures in between those of G-type stars and M dwarfs. Stars with similar age as Kepler-444\,A (about 10\,Gyr) can display Ly-$\alpha$ line profiles that are either SP (HD\,97658, \citealt{Bourrier2016_HD976}) or DP (HD\,166, \citealt{Wood2005}), but the fact that about three quarters of the sampled K dwarfs show a self-reversal favors a similar shape for Kepler-444\,A.\\
We conclude that the intrinsic Lyman-$\alpha$ line of Kepler-444\,A is more likely to be double-peaked, and we used this line profile in the rest of our study. It is displayed in Fig.~\ref{fig:theo_spectrum}, with the parameters derived for the ISM column density and Ly-$\alpha$ integrated flux given in Table~\ref{tab:results_ISM_La}.\\

\subsection{XUV emission}
\label{sec:xuv}
 
We estimated the XEUV spectrum of Kepler-444\,A from 5 to 912\,\AA\, to infer the photoionization rate of neutral hydrogen in the system, and to predict the energy-limited mass loss rates of the outer planets. \\
The EUV spectrum is mostly absorbed by the ISM and was calculated with the semi-empirical relations from \citet{Linsky2014}, which are based on the intrinsic Ly-$\alpha$ flux. We give in Table~\ref{tab:results_ISM_La} the flux integrated in the 100--912\,\AA\, band, with uncertainties accounting for both uncertainties on the Ly-$\alpha$ flux, and on the empirical relations themselves. We compare the derived EUV flux with that predicted for the Sun at an age of 11.2\,Gyr (\citealt{Ribas2005}). Using their relations in the bands 100--360\,\AA\, and 360--920\,\AA\,, we found F$_\mathrm{EUV}^\mathrm{Sun}\sim$1.2\,erg\,cm$^{-2}$\,s$^{-1}$ at 1\,au. We also estimated the future solar Ly-$\alpha$ flux and found F$_{\mathrm{Ly}\alpha}^\mathrm{Sun}$=3.4\,erg\,cm$^{-2}$\,s$^{-1}$ at 1\,au. The predicted solar EUV and Ly-$\alpha$ fluxes are thus very close to the values derived from our reconstruction of Kepler-444\,A Ly-$\alpha$ line profile in Sect.~\ref{sec:Lalpha_rec}, giving us further confidence that it is double-peaked. \\

The X-ray emission of Kepler-444\,A was derived at 1\,au from the star using several empirical relations. \citet{Wood2005} derived a relation for K\,V stars based on the Ly-$\alpha$ flux, which yields F$_\mathrm{X}\sim$0.04\,erg\,cm$^{-2}$\,s$^{-1}$ in 5--100\,\AA\,. The relation for K0\,V-K5\,V stars in \citet{Linsky2013} yields a similar flux of $\sim$0.01\,erg\,cm$^{-2}$\,s$^{-1}$. We then used the EUV fluxes derived from our Ly-$\alpha$ reconstruction to estimate the X-ray flux in the band 5.2--124\,\AA\,. The \citet{Chadney2015} relation yields a larger flux of about 0.16\,erg\,cm$^{-2}$\,s$^{-1}$. Similarly, the age-luminosity relation in \citet{SanzForcada2010} (see also \citealt{Garces2011}) yields F$_\mathrm{X}$=0.16\,erg\,cm$^{-2}$\,s$^{-1}$, while the relation in \citet{Jackson2012}, based on a larger sample, yields F$_\mathrm{X}$=0.21\,erg\,cm$^{-2}$\,s$^{-1}$. The differences in those X-ray fluxes might be linked to the different spectral bands used by these studies, the large uncertainties attached to the empirical relations, or the fact that they are all based on stars younger than Kepler-444\,A. Direct measurements of the X-ray spectrum of Kepler-444\,A will be required to determine precisely its high-energy emission, but hereafter we consider that it can be approximated by a flux of 0.1\,erg\,cm$^{-2}$\,s$^{-1}$. It is noteworthy that the active K1 host-star HD\,97658, with an age close to that of Kepler-444\,A (9.7$\pm$2.8\,Gy), was found to emit fluxes at 1\,au of about 8\,erg\,cm$^{-2}$\,s$^{-1}$ at Ly-$\alpha$ and 0.7\,erg\,cm$^{-2}$\,s$^{-1}$ in the X-ray, larger than for Kepler-444\,A. This might be another indication that Kepler-444\,A has not a strong chromospheric activity (Sec.\ref{sec:st_var}).


\section{Evaporating sub-Earth planets}
\label{sec:evap_planets}

\subsection{Estimations of present escape}
\label{sec:esc_rate}

We proposed in Sect.~\ref{sec:pl_abs} that the variations observed in the Ly-$\alpha$ line could arise from hydrogen exospheres around the outer planets Kepler-444\,e and f. Their small radii imply that they do not possess hydrogen-dominated atmospheres. Planetary thermal evolution models (Lopez \& Fortney 2014) also predict that the composition of planets with radii less than 0.8\,R$_\mathrm{Earth}$ are likely rocky. Yet in Dec. 2016 less than ten exoplanets in this radius range had known masses\footnote{From the Extrasolar Planets Encyclopaedia at exoplanet.eu, \citet{Schneider2011}}, and there is thus scant observational constraints on the formation and evolution of sub-Earth size planets. The study of the Kepler-444 system by \citet{Dupuy2016} suggests that the disk in which formed the planets, truncated by Kepler-444\,BC at 1-2\,au, would have been depleted in volatiles, leading to small, high-density planets. However, the proximity of each planet to a strong, first-order mean-motion resonance indicates that the system evolved dynamically after the formation of the planets (\citealt{Campante2015}). Such a configuration could arise from convergent inward migration, and here we consider a scenario where the ice line was within the truncation radius of the protoplanetary disk. If the planets formed beyond the ice line before migrating, they could have accreted a large fraction of water ice. Assuming a Ganymede-like albedo, the equilibrium temperature of the planets would range between about 640 and 890\,K, larger than water critical temperature (647\,K). Stellar irradiation could thus be sufficient for the surface ice to melt and form a supercritical steam atmosphere (\citealt{Jura2004}, \citealt{Howe2014}). We calculated the total mass-loss rate from planets e and f in the energy-limited regime (\citealt{Lecav2007}):
\begin{equation}
\label{eq:H_esc_rate}
\dot{M}^{tot}= \eta \, (\frac{R_\mathrm{XUV}}{R_\mathrm{p}})^2 \, \frac{3 \, F_\mathrm{XUV}(\mathrm{sma})}{4 \, G \, \rho_\mathrm{p} \, K_{tide}}.  
\end{equation}
where $(\frac{R_\mathrm{XUV}}{R_\mathrm{p}})^2$ accounts for the increased cross-sectional area of planets to EUV radiation, and $K_\mathrm{tide}$ accounts for the contribution of tidal forces to the potential energy (\citealt{Erkaev2007}). Both are set to unity for these very small planets. The heating efficiency $\eta$ is the fraction of the stellar energy input that is available for atmospheric heating (e.g., \citealt{Ehrenreich_desert2011}, \citealt{Lammer2013}; \citealt{Shematovich2014}). We assumed a Ganymede-like composition, with mean density 1.94\,g\,cm$^{-3}$ and half of the planet mass made of water ice.\\
Mass loss rates were calculated with the X-ray and EUV fluxes given in Table~\ref{tab:results_ISM_La}. We found $\dot{\mathrm{M}}^{tot}_\mathrm{e}$ = $\eta$\,1.7$\times$10$^{9}$\,g\,s$^{-1}$ and $\dot{\mathrm{M}}^{tot}_\mathrm{f}$ = $\eta$\,1.2$\times$10$^{9}$\,g\,s$^{-1}$. In that case the age of the system puts upper limit on $\eta$ at 40\% and 130\% for planets \textit{e} and \textit{f}, which means that even with conservative efficiencies of 10\% (\citealt{Ehrenreich2015}, \citealt{Salz2016a}) those planets would only have lost about 27\% and 8\% of their water content, respectively, after 11.2\,Gy. Even though those results do not account for the temporal evolution of the stellar flux, and use simple assumption on the complex physics behind atmospheric escape, they leave open the possibility that the old planets of the Kepler-444 system could still be ``hot Ganymedes'' with a high water content.\\

\subsection{Simulations of hydrogen exospheres}
\label{sec:sim_EVE}

We used the EVaporating Exoplanets (EVE) code (\citealt{Bourrier_lecav2013}, \citealt{Bourrier2015_GJ436}) to interpret the Ly-$\alpha$ variations in terms of exospheric absorption. Given that this is a tentative interpretation, and that the available data have a low SNR, we did not perform a full exploration of the parameter space. Rather, we searched if there was a configuration that could reproduce the observations given the physical conditions in the planetary system.

We assumed circular orbits for the planets, as their eccentricity was found to be consistent with 0 \citep{Campante2015,Hadden2016}. We assumed Ganymede-like compositions to estimate the planets' masses from their measured radius. Hydrogen was launched from half the planets Roche lobe radius, with an upward velocity of 5\,km\,s$^{-1}$. This velocity was found to yield good fits to the data, and is in the range predicted by \citet{Salz2016b} for small planets with low gravitational potential. We note that the launch conditions have a second-order influence on the fit compared to the escape rate $\dot{\mathrm{M}}_{\mathrm{H^{0}}}$. The dynamics of the escaping particles depends on the velocity-dependent stellar radiation pressure, derived from the Ly-$\alpha$ line reconstruction in Sect.~\ref{sec:Lalpha_rec}. Neutral hydrogen particles are also subjected to stellar photoionization, with a rate $\Gamma_{\mathrm{ion}}$ calculated using the XUV spectrum derived in Sect.~\ref{sec:xuv}. Theoretical Ly-$\alpha$ spectra generated with the EVE code were compared to the observations between -250\,km\,s$^{-1}$ and the limit of the airglow range (Fig.~\ref{fig:grid_sp_fits})

\subsubsection{Kepler-444 e and f}
\label{sec:fit_K444ef}

Based on the discussion in Sect.~\ref{sec:pl_abs}, Kepler-444\,e and f are the most likely candidates to explain the variations observed in Visits A, B, C, and F. We simulated the transits of their exospheres in those four visits, phased over Kepler-444f ephemeris in Visit A (Fig.~\ref{fig:light_curve_plf}). This allowed us to simulate the exospheres during the same orbital revolution of the planets. Spectra were averaged over their common phase windows during the transit of planet f, so as to mitigate the effects of stellar variability or the additional absorption by other planet's exospheres in each visit. \\

\begin{figure*}     

\includegraphics[trim=0.9cm 0cm 1.5cm 0cm,clip=true,width=1.5\columnwidth]{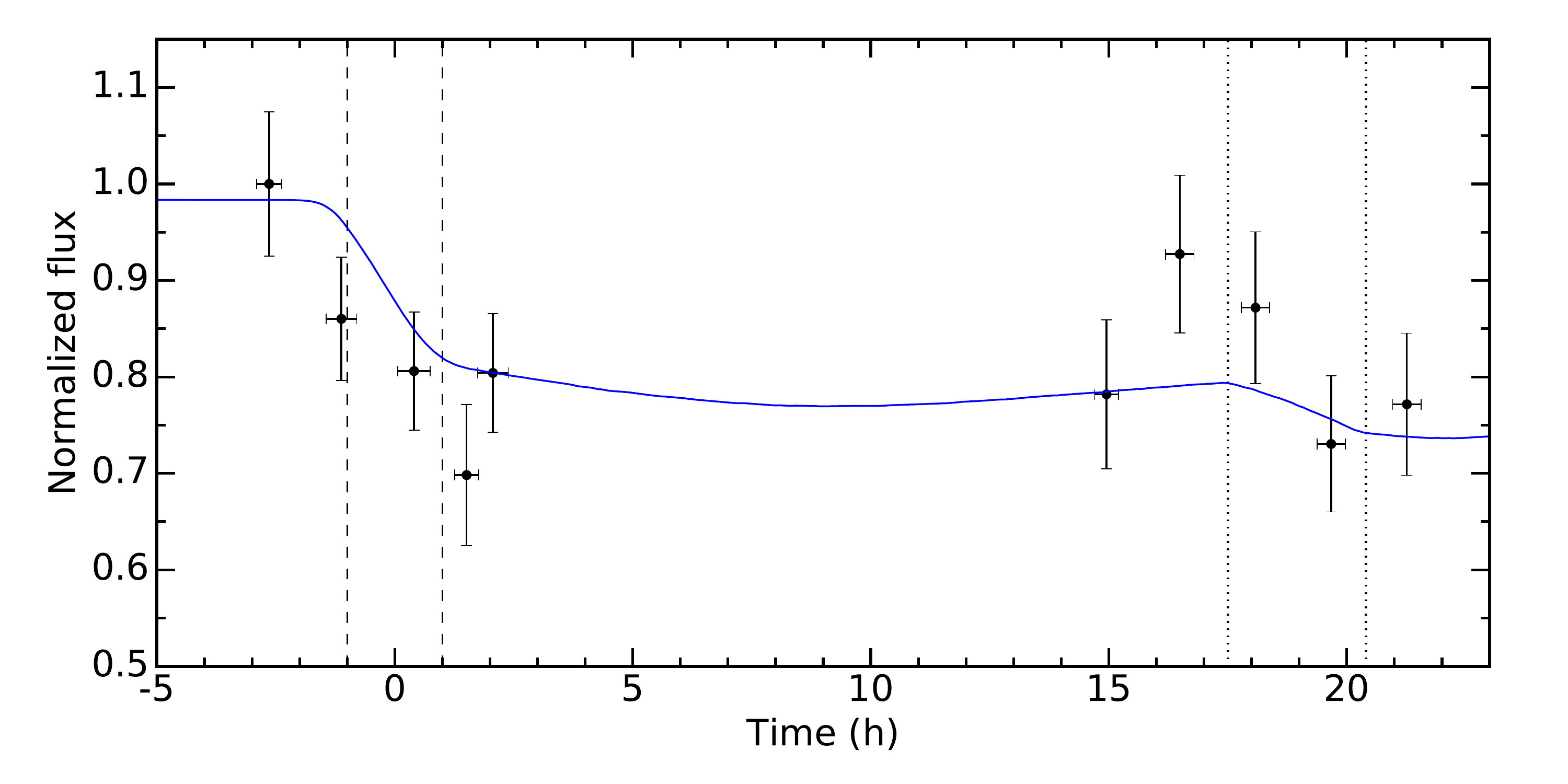}
\includegraphics[trim=0.7cm 0cm 0.cm 0cm,clip=true,width=0.397\columnwidth]{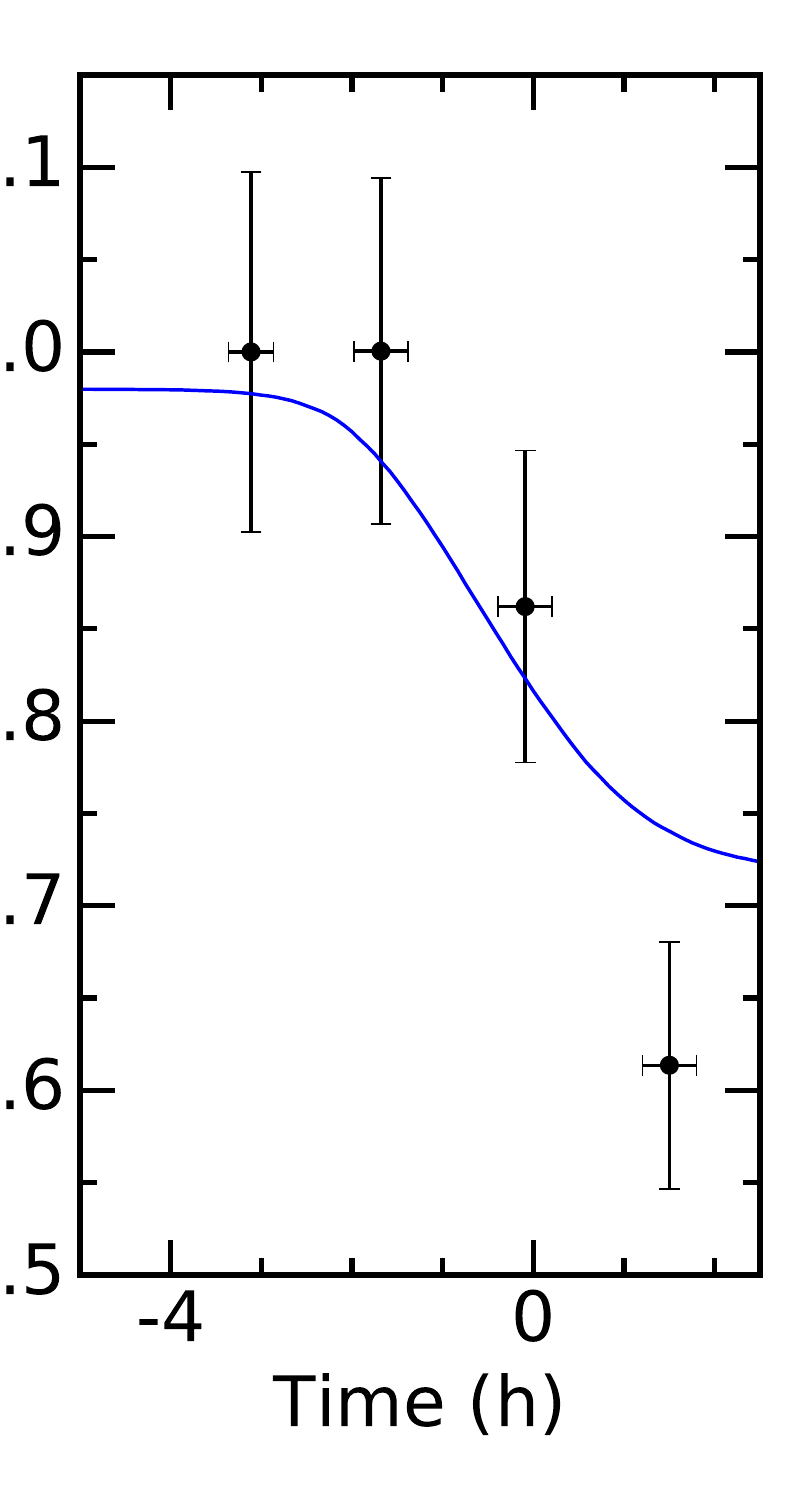}
\caption[]{Transit light curve in the Ly-$\alpha$ line, for Kepler-444\,e+f (left panel) and Kepler-444\,g (right panel). In the left panel, spectra in Visits A, B, C, and F have been averaged over common phase windows after being phased over Kepler-444\,f ephemeris, and integrated in the range [-89;-37]\,km\,s$^{-1}$ (see first and second column in Fig.~\ref{fig:grid_sp_fits}). In the right panel, spectra in Visit E have been integrated in the range [-103;-37]\,km\,s$^{-1}$  (see third column in Fig.~\ref{fig:grid_sp_fits}). In blue are plotted the theoretical transit light curves of the planets' exospheres, subjected to a low photoionization and radiation pressure, and to interactions with the stellar wind. They were fitted independently to the data in the left and right panel.}
\label{fig:LC_fits}
\end{figure*}

Because Ly-$\alpha$ radiation pressure from Kepler-444\,A barely compensates for its gravity (Fig.~\ref{fig:theo_spectrum}), we found that it could not explain the acceleration of the escaping hydrogen to the observed velocities. We thus investigated the additional effect of charge-exchange between the planetary exosphere and the stellar wind (\citealt{Holmstrom2008}, \citealt{Ekenback2010}, \citealt{Bourrier_lecav2013}). Stellar wind interactions typically abrade the regions of the exosphere facing the star, and create a secondary tail of neutralized protons with the dynamics of the stellar wind. Similarly to the case of the warm Neptune GJ\,436b (\citealt{Ehrenreich2015}, \citealt{Bourrier2016}) we found that the combination of stellar wind interactions, radiative braking, and low photoionization rates (neutral hydrogen atoms have photoionization lifetimes of about 19\,h at planet e and 26\,h at planet f) could reproduce reasonably well the observations (Fig.~\ref{fig:grid_sp_fits}). The light curve in Fig.~\ref{fig:LC_fits} illustrates how the cloud of stellar wind protons, neutralized by charge exchange with planet \textit{f} exosphere, is extended enough to keep occulting the star up to the time of  planet \textit{e} transit. Observations are well fitted with mass loss rates in the range $\dot{M}(\mathrm{H^{0}})_{\mathrm{e}}\sim$[10$^{7}$ ; 10$^{8}$]\,g\,s$^{-1}$ and $\dot{M}(\mathrm{H^{0}})_{\mathrm{f}}\sim$[4$\times$10$^{7}$ ; 10$^{8}$]\,g\,s$^{-1}$, with a density of stellar wind protons of about 4$\times$10$^{7}$\,m$^{-3}$ (at 1\,au from the star) and a bulk velocity of $\sim$60\,km\,s$^{-1}$. We note that this velocity is about four times lower than that of the very slow solar wind (e.g., \citealt{SanchezDiaz2016}), or the wind of the K-type host star HD\,189733b (as derived by \citealt{Bourrier_lecav2013} from transit observations of its evaporating hot Jupiter companion). However, wind velocity is expected to decrease with age and Kepler-444\,A is more than twice older than these stars (see \citealt{Boyajian2015} for HD\,189733). We can also note that \citet{Bourrier2016} derived a low velocity of about 85\,km\,s$^{-1}$ for the stellar wind of the M dwarf GJ\,436, which interacts with the giant exosphere of its evaporating warm Neptune companion. We note that the derived mass loss rates of neutral hydrogen are consistent with the larger total mass loss rates estimated in Sect.~\ref{sec:esc_rate}, and would suggest a low heating efficiency and/or a low neutral hydrogen fraction. We remind the reader that this is a tentative interpretation of the data, and that a better understanding of the physical conditions in the Kepler-444 system will require further observations.\\

\subsubsection{A sixth planet in the system?}
\label{sec:444g}

In Sect.~\ref{sec:pl_abs} we reported a significant flux decrease in the Ly-$\alpha$ line during Visit E. Using the out-of-transits spectrum defined in Sect.~\ref{sec:pl_abs} as reference for the unocculted stellar line, this variation corresponds to an absorption of the stellar flux by 40$\pm$6\% (6.5\,$\sigma$) in the peak band (Fig.~\ref{fig:grid_sp_fits}). We suggest that this variation could arise from a sixth planet in the Kepler-444 system, whose exosphere would graze the stellar disk. This situation is reminiscent of the warm Jupiter 55 Cnc b, whose planetary disk does not transit the star at optical wavelength but which was observed in transit at Ly-$\alpha$ because of its extended hydrogen exosphere (\citealt{Ehrenreich2012}). To produce such a deep transit at Ly-$\alpha$, an evaporating Kepler-444\,g would likely have an impact parameter $\approxinf$1.1. Assuming that the system is coplanar, a value of $b$=1.1 corresponds to an orbital distance of $\sim$0.13\,au (P$\sim$19\,d), which is close to the predicted limit of 0.15\,au where the atmosphere of a giant planet around a solar-like star is expected to retrieve its stability (\citealt{Koskinen2007}). Given that the radii of the five known Kepler-444 planets increase monotonically outwards, from Mars to Venus size, we assumed the hypothetical Kepler-444\,g is Earth-size with the composition of Ganymede. We then compared the signature of its theoretical exosphere to the spectra in Visit E (Fig.~\ref{fig:grid_sp_fits}), assuming that it interacts with the same stellar wind and radiation pressure as Kepler-444\,e and f in Sect.~\ref{sec:fit_K444ef}. In these conditions, an exospheric transit could explain well the observed variation if Kepler-444\,g had its inferior conjunction at BJD$\sim$2457546 and an escape rate $\dot{M}(\mathrm{H^{0}})_{\mathrm{g}}\sim$1.5$\times$10$^{8}$\,g\,s$^{-1}$. To yield such a deep absorption shortly after ingress, Kepler-444\,g would have to be surrounded by a giant exosphere similar to that of the warm Neptune GJ\,436b (\citealt{Ehrenreich2015}, \citealt{Bourrier2015_GJ436}). The presence of absorption at high velocities in the peak band of Visit E, but not at lower velocities in the core band (see Fig.~\ref{fig:light_curves} and Fig.~\ref{fig:grid_sp_fits}), also bears a strong resemblance with observations of the evaporating planets GJ\,436b and HD\,189733b (\citealt{Lecav2012}, \citealt{Bourrier2013}). In both cases  observations were well explained by charge-exchange with the stellar wind (\citealt{Bourrier_lecav2013}, \citealt{Bourrier2016}), which ionize the low-velocity neutral hydrogen close to the planet and replace it with a population of neutralized protons moving with the higher velocity of the stellar wind. Such interactions also increase the spatial extension of the exospheric tail, which suggests that observations scheduled up to more than 10\,hours after the predicted inferior conjunction of Kepler-444\,g could still probe the transit of its exosphere, allowing us to confirm the existence of a sixth planet in the Kepler-444 system.

\section{Conclusion}
\label{sec:conclu}

With its blueshifted radial velocity, the Kepler-444 system is a compelling target for exospheric characterisation of exoplanets smaller than the Earth. Kepler-444\,A hosts five planets ranging in size between Mars and Venus, in a system 2.5 times older than the Solar system. The planets are too small today to possess large hydrogen-rich envelopes, which would probably have been lost a long time ago through atmospheric escape. However, the planets could have migrated from beyond the ice line after gathering large amounts of water ice. Planets with Ganymede-like compositions orbiting closer than about 0.08\,au from their K-type star could possibly be enshrouded in a thick envelope of supercritical steam (\citealt{Jura2004}). This water envelope could then provide a large source of atomic hydrogen under the photodissociated effect of the incoming stellar EUV and X-ray radiation.\\
We used the HST/STIS to search for the signature of hydrogen escape from such envelopes, observing the Ly-$\alpha$ line of Kepler-444\,A at six independent epochs. We detected significant flux variations during the transits of Kepler-444\,e (one epoch, $\sim$20\%), Kepler 444\,f (three epochs, $\sim$20\%), and surprisingly at a time when none of the known planets was transiting (one epoch, $\sim$40\%). While variability in the transition region and corona of the host star is a possible explanation for these variations, we show that they can also be attributed to the transits of neutral hydrogen exospheres trailing the two outer planets Kepler 444\,e and \,f, and the partial transit of an exosphere from an undetected planet Kepler-444\,g. \\
Observations of the sodium doublet with HARPS-N allowed us to identify and characterize two ISM clouds along the line-of-sight toward Kepler-444\,A. This allowed us to reconstruct the intrinsic profile of the stellar Ly-$\alpha$ line and to estimate the XUV emission of the star, which would still allow for a moderate mass loss from the outer planets after 11.2\,Gy. Performing simulations with the EVE code, we found that the observations could then be explained with neutral hydrogen mass loss rates in the order of 10$^{7}$ -- 10$^{8}$\,g\,s$^{-1}$ for Kepler-444\,e, \,f, and \,g, assuming that their exospheres are shaped by both radiative braking and wind from the host star. \\
More observations of the Kepler-444 system at Ly-$\alpha$ and in the X-rays will be necessary to assess the variability of the host star, and search for a sixth planet in the system. The possible detection of hydrogen exospheres around the outer planets hints at the tantalizing possibility that they are hot Ganymedes with a large water content. This would shed a new light on our understanding of planetary formation and internal structure models of exoplanets smaller than the Earth. \\


\begin{acknowledgements}
We thank the referee, Jeffrey Linsky, for his valuable comments that helped improve our study. V.B. warmheartedly thanks G. F\'eraud for coming up with the idea of a sixth planet. The authors warmly thanks A. Verhamme, P. Eggenberger, and M. Audard for their help with the Ly-$\alpha$ line and properties of Kepler-444\,A. This work is based on observations made with the NASA/ESA Hubble Space Telescope, obtained at the Space Telescope Science Institute, which is operated by the Association of Universities for Research in Astronomy, Inc., under NASA contract NAS 5-26555. This work has been carried out in the frame of the National Centre for Competence in Research ``PlanetS'' supported by the Swiss National Science Foundation (SNSF). V.B. acknowledges the financial support of the SNSF. This research has made use of the Extrasolar Planets Encyclopaedia at exoplanet.eu.
\end{acknowledgements}

\bibliographystyle{aa} 
\bibliography{biblio} 

\begin{appendix}

\section{Appendix}
\label{apn:mcmc}

\begin{figure*}
\centering
\begin{minipage}[b]{\textwidth}   
\includegraphics[trim=1.5cm 5.1cm 10cm 3.3cm,clip=true,width=0.384\textwidth]{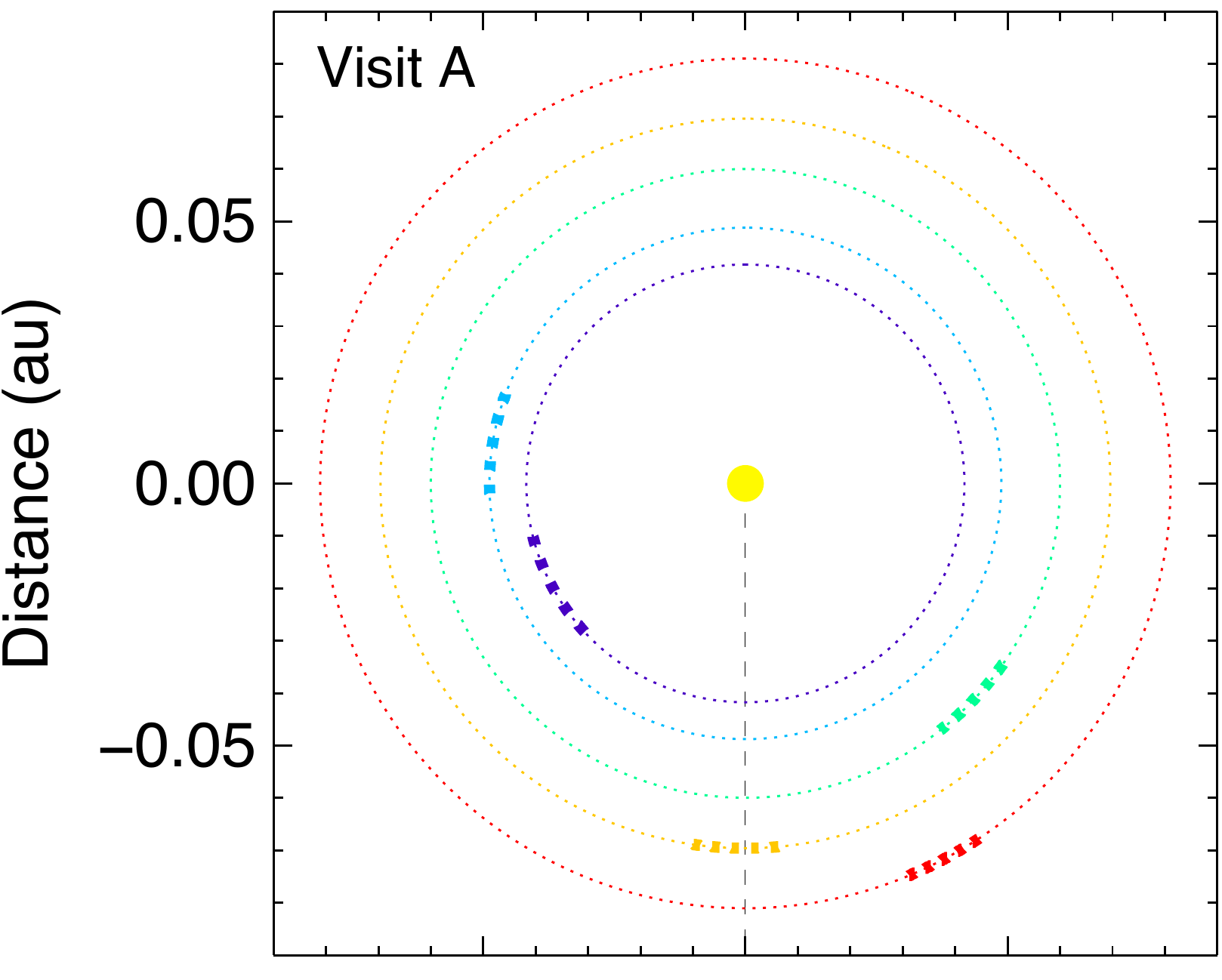}
\includegraphics[trim=5.1cm 5.1cm 10cm 3.3cm,clip=true,width=0.3\textwidth]{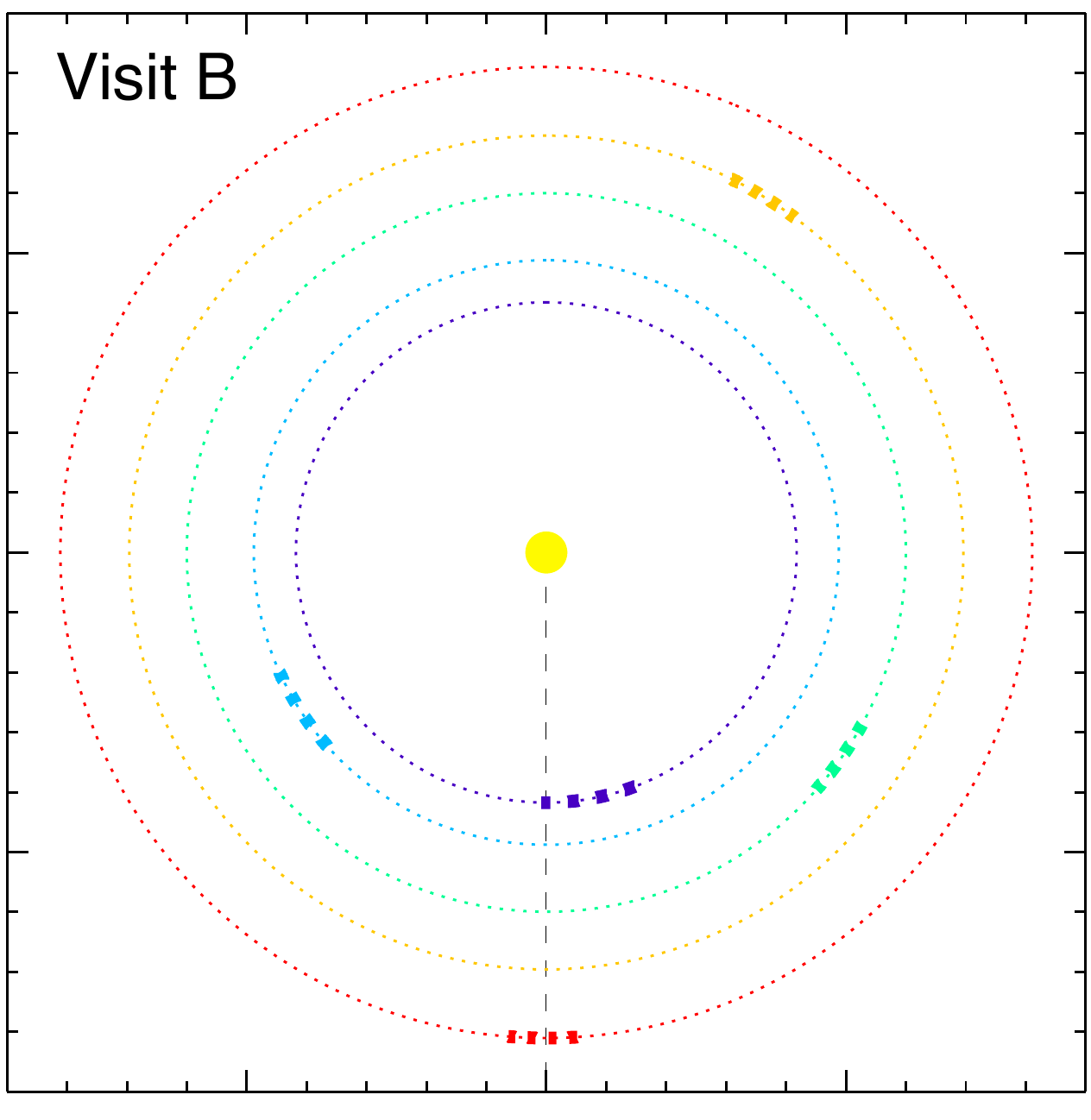}
\includegraphics[trim=5.1cm 5.1cm 10cm 3.3cm,clip=true,width=0.3\textwidth]{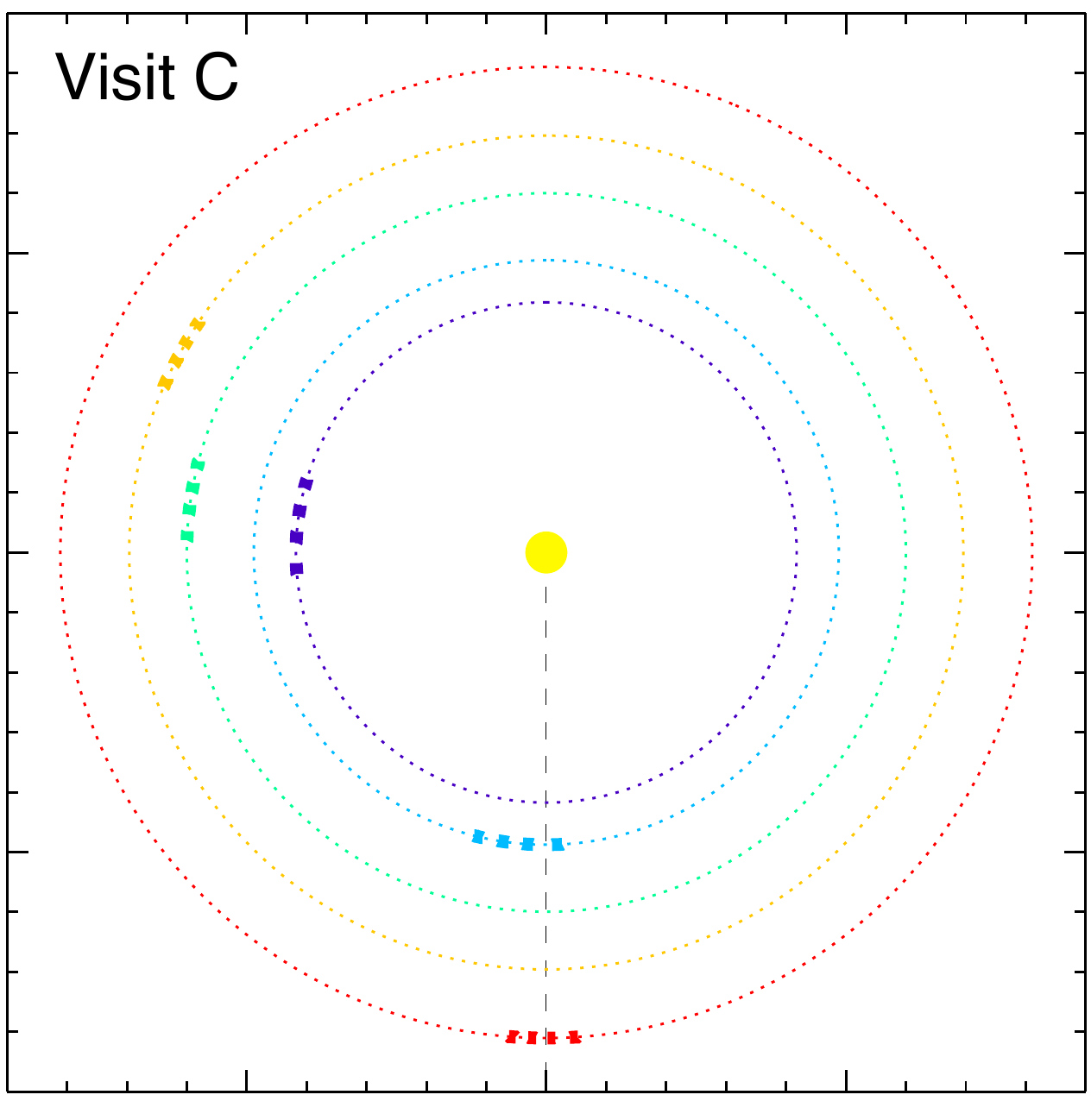}\\
\includegraphics[trim=1.5cm 3.1cm 10cm 3.3cm,clip=true,width=0.384\textwidth]{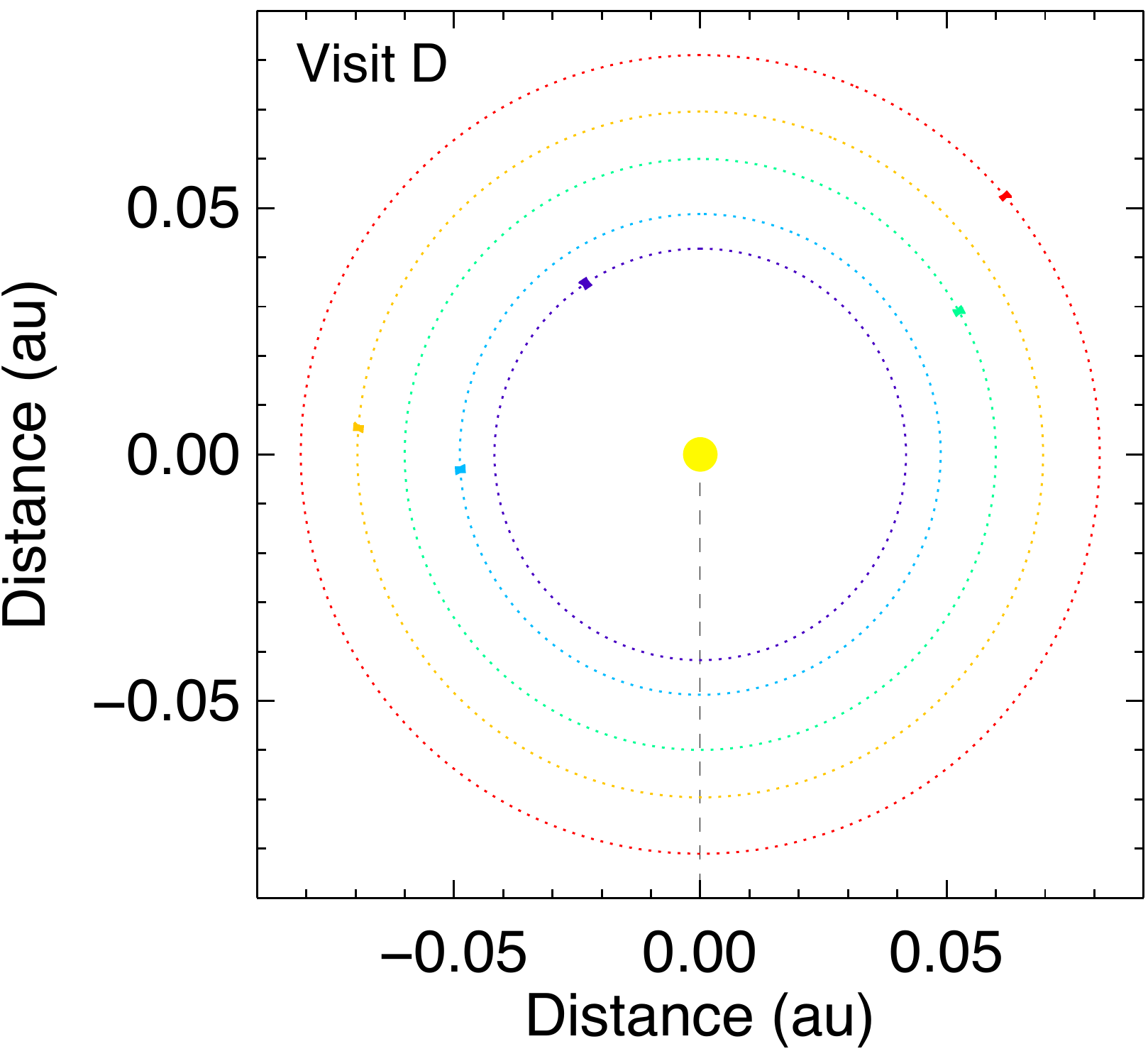}
\includegraphics[trim=5.1cm 3.1cm 10cm 3.3cm,clip=true,width=0.3\textwidth]{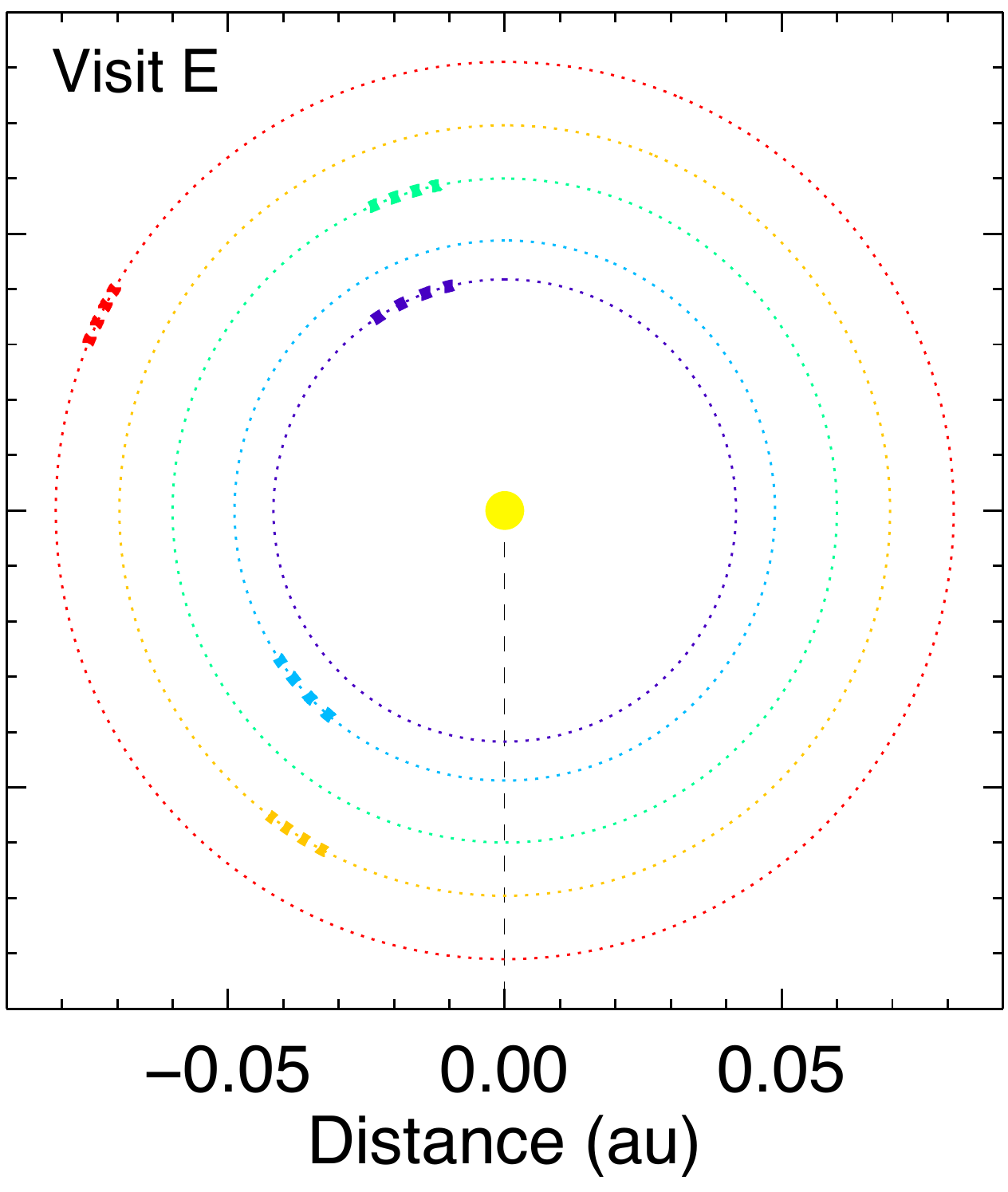}
\includegraphics[trim=5.1cm 3.1cm 10cm 3.3cm,clip=true,width=0.3\textwidth]{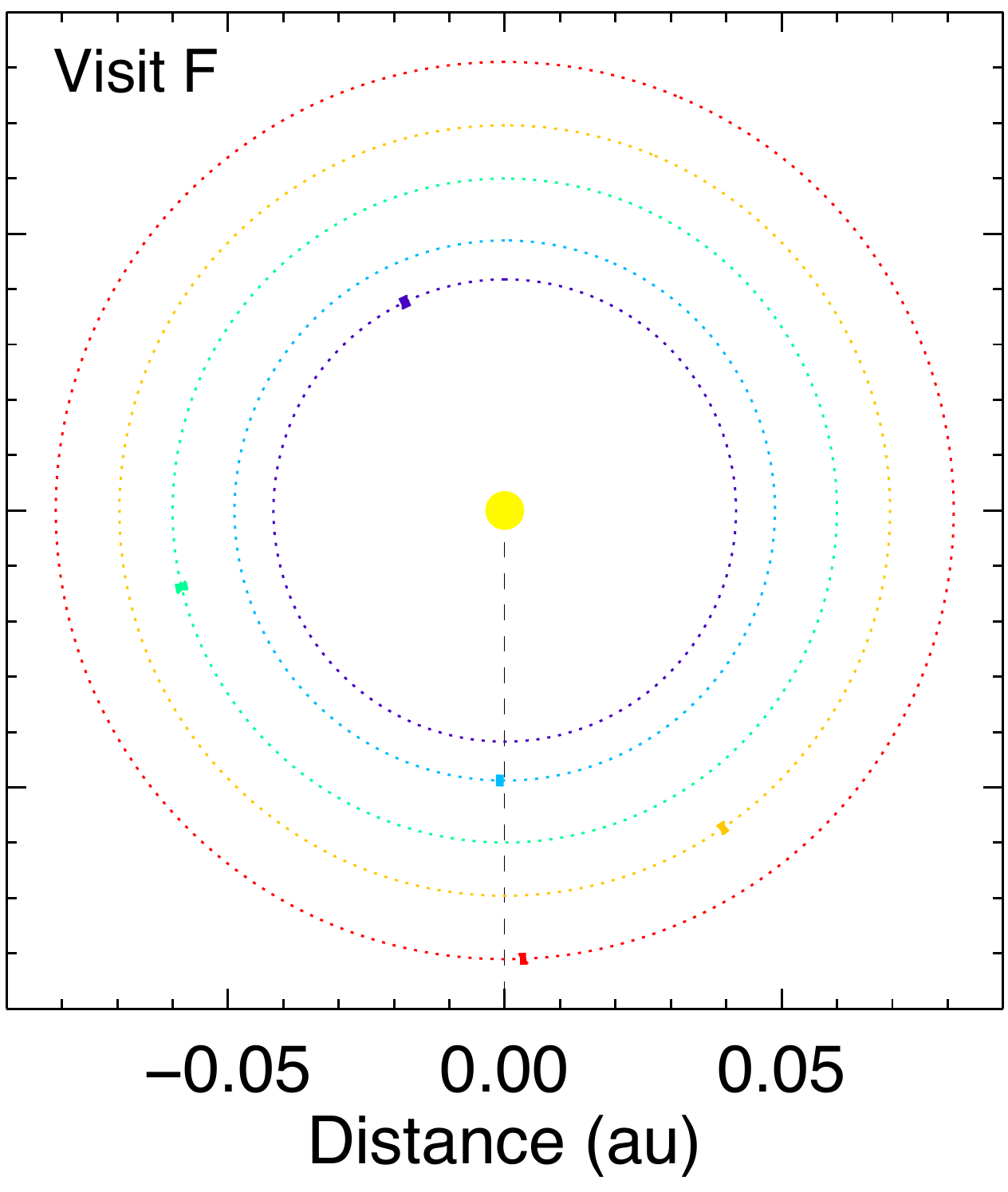}\\
\end{minipage}
\caption[]{Orbital positions of the Kepler-444 planets at the time of the HST observations. Each rectangle corresponds to the space covered by a planet during one of the HST orbits. Planets are moving counterclockwise. The dashed black line indicates the LOS toward Earth. Star and orbital trajectory have the correct relative scale.}
\label{fig:orb_cov}
\end{figure*}

\begin{figure*}
\centering

\begin{minipage}[b]{\textwidth}   
\includegraphics[trim=1.2cm 7.8cm 1.35cm 8.5cm,clip=true,width=0.3565\textwidth]{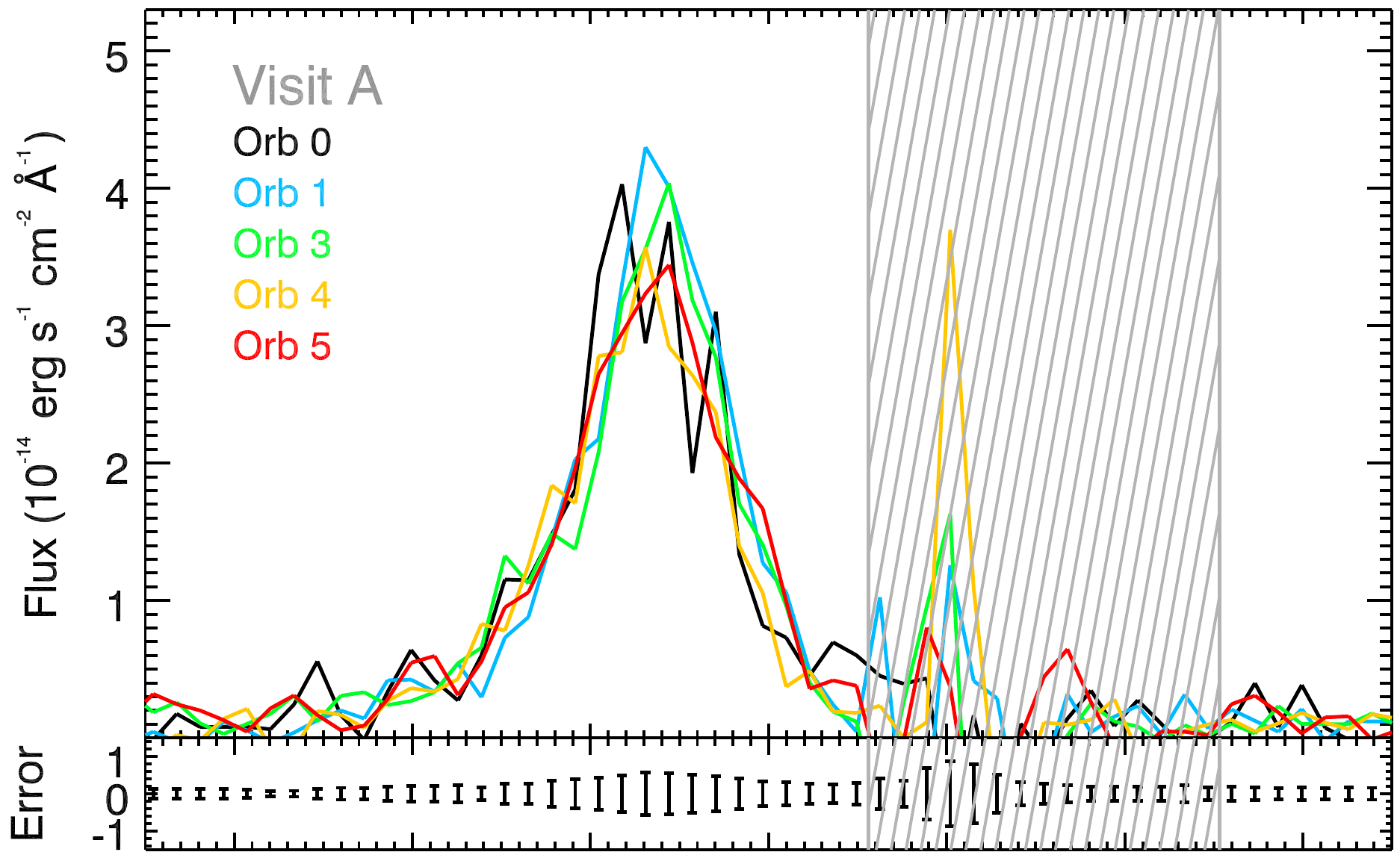}
\includegraphics[trim=3.1cm 7.8cm 1.35cm 8.5cm,clip=true,width=0.321\textwidth]{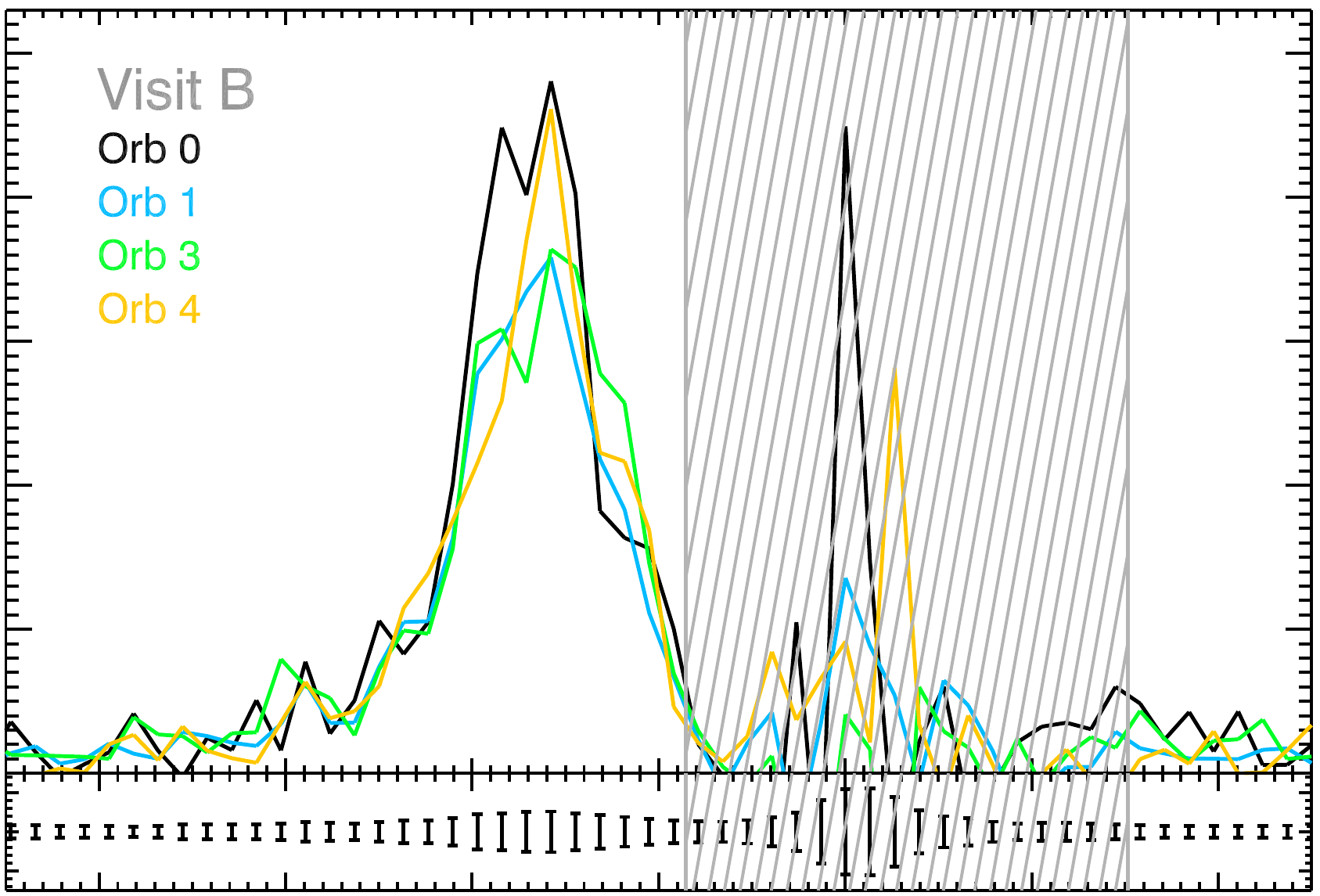}
\includegraphics[trim=3.1cm 7.8cm 1.35cm 8.5cm,clip=true,width=0.321\textwidth]{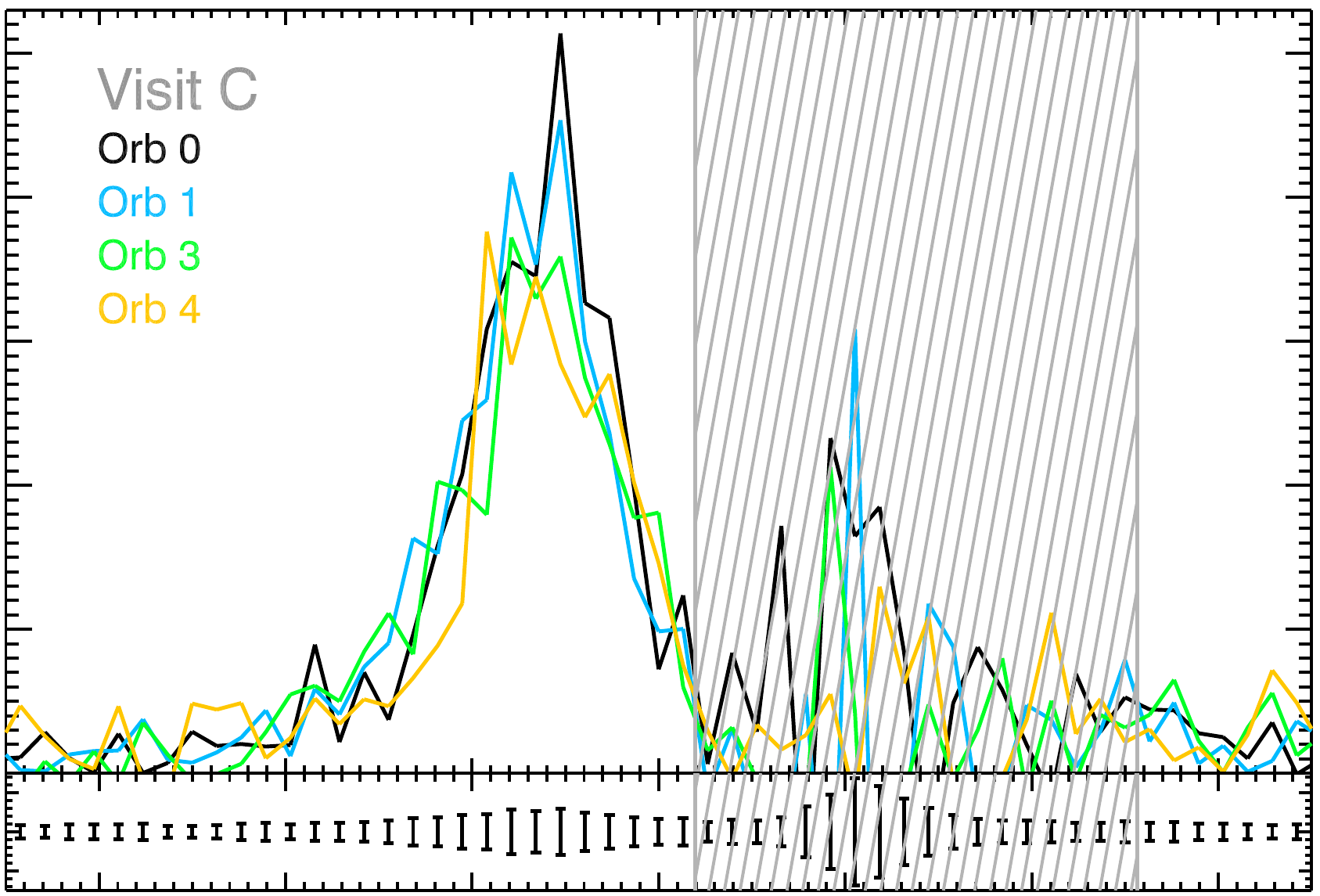}\\
\includegraphics[trim=1.2cm 5.5cm 1.35cm 8.5cm,clip=true,width=0.3565\textwidth]{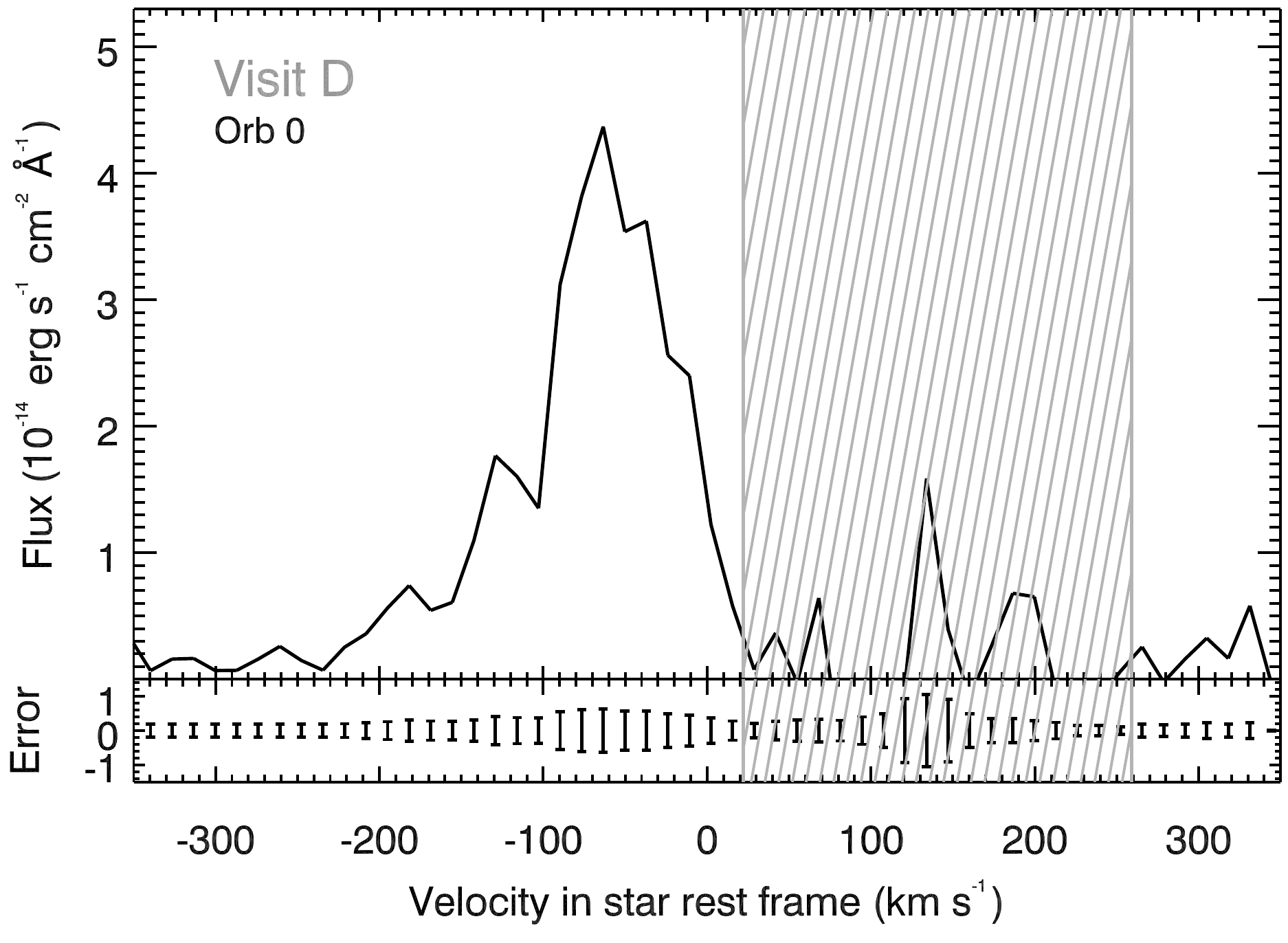}
\includegraphics[trim=3.1cm 5.5cm 1.35cm 8.5cm,clip=true,width=0.321\textwidth]{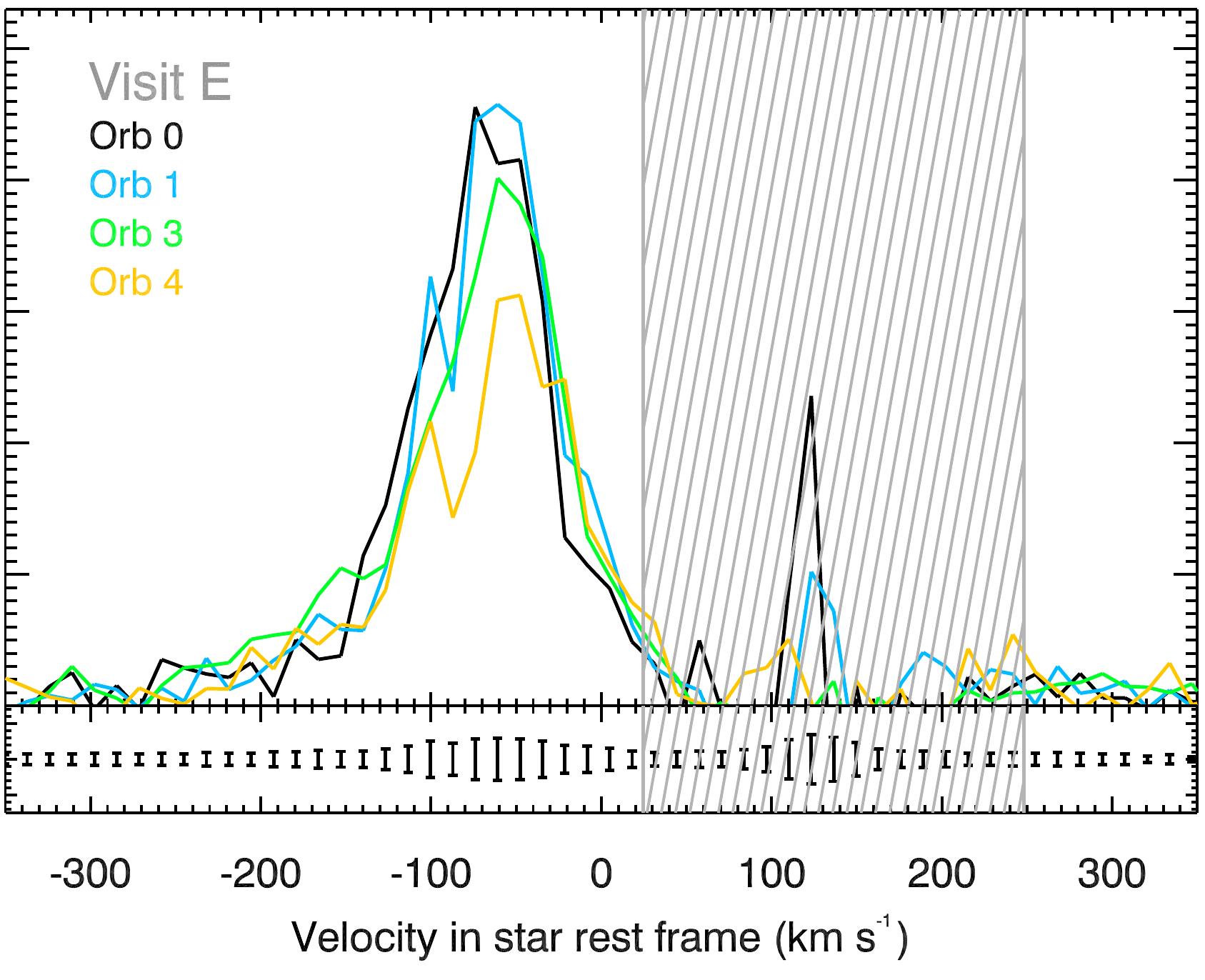}
\includegraphics[trim=3.1cm 5.5cm 1.35cm 8.5cm,clip=true,width=0.321\textwidth]{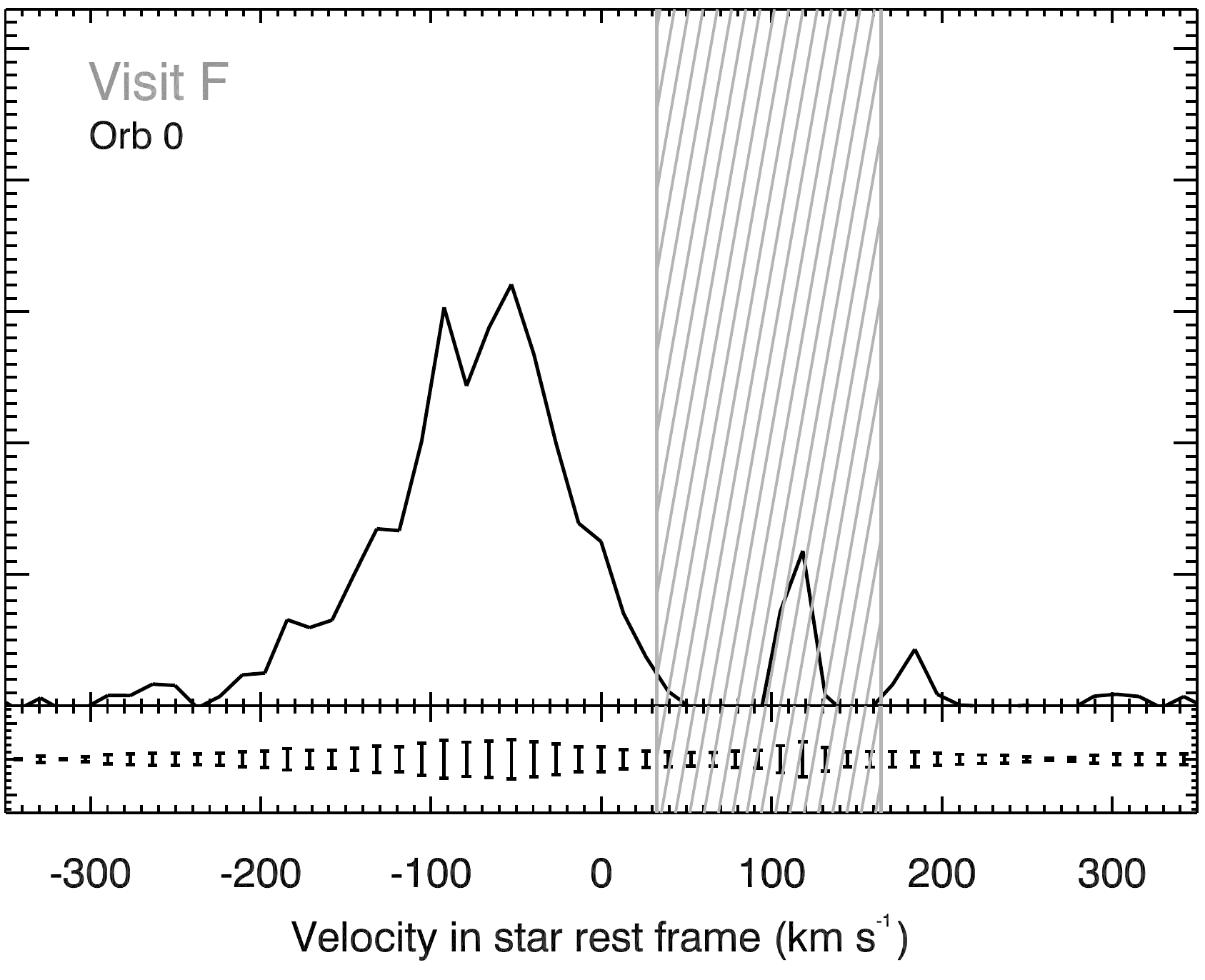}\\
\end{minipage}

\caption[]{Measurements of Kepler-444 Ly-$\alpha$ line during each HST visit. We show the spectra gathered over the full HST orbits, and plotted as a function of Doppler velocity in the stellar rest frame. Subpanels display error bars typical of a full-orbit spectrum in each visit. The shaded gray area corresponds to the range affected by ISM absorption and geocoronal emission.}
\label{fig:grid_spec}
\end{figure*}

\begin{figure*}
\centering
\begin{minipage}[b]{\textwidth}   
\includegraphics[trim=0cm 2.2cm 1cm 0cm,clip=true,width=0.3535\textwidth]{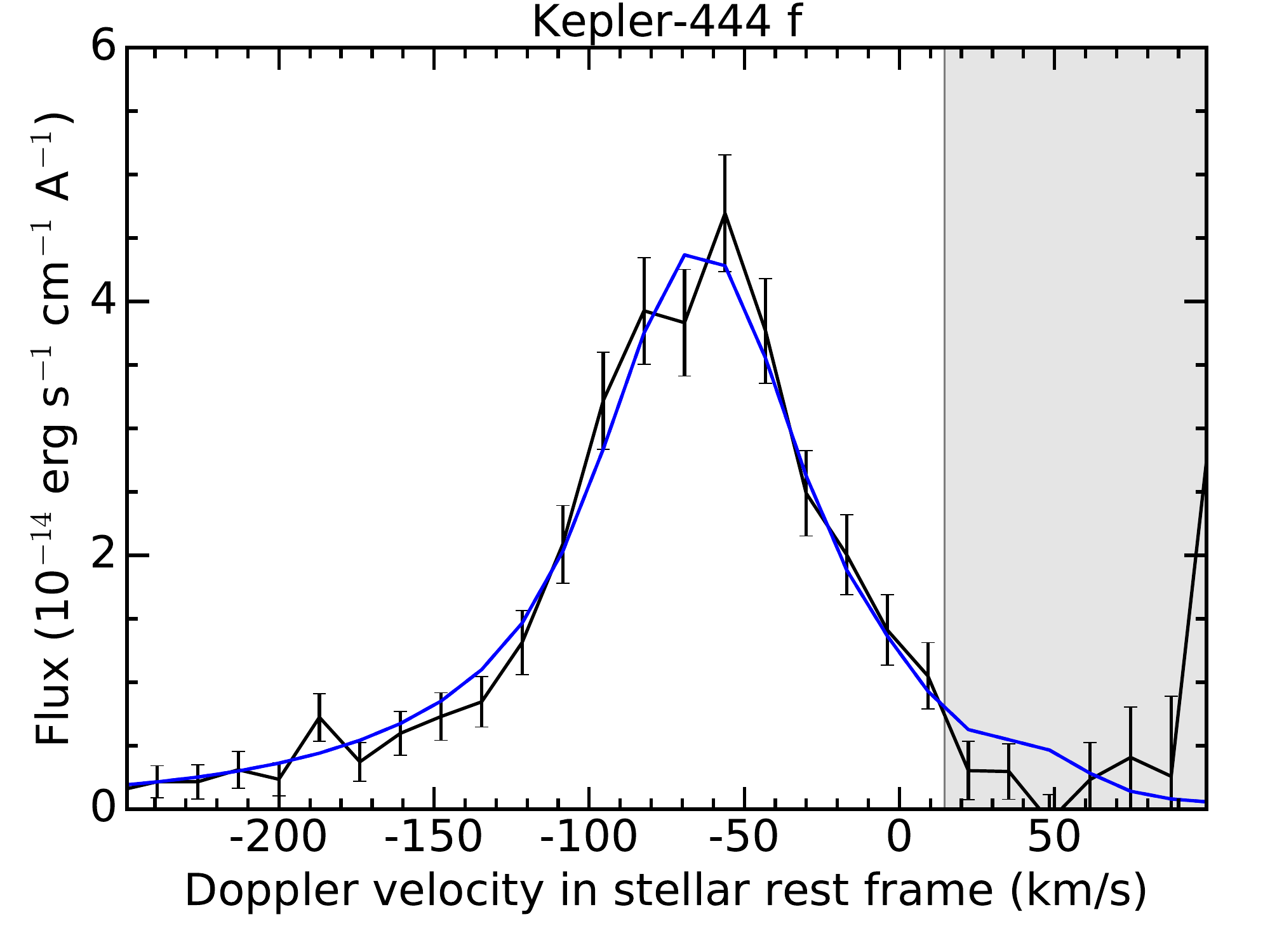}
\includegraphics[trim=2cm 2.2cm 1cm 0cm,clip=true,width=0.317\textwidth]{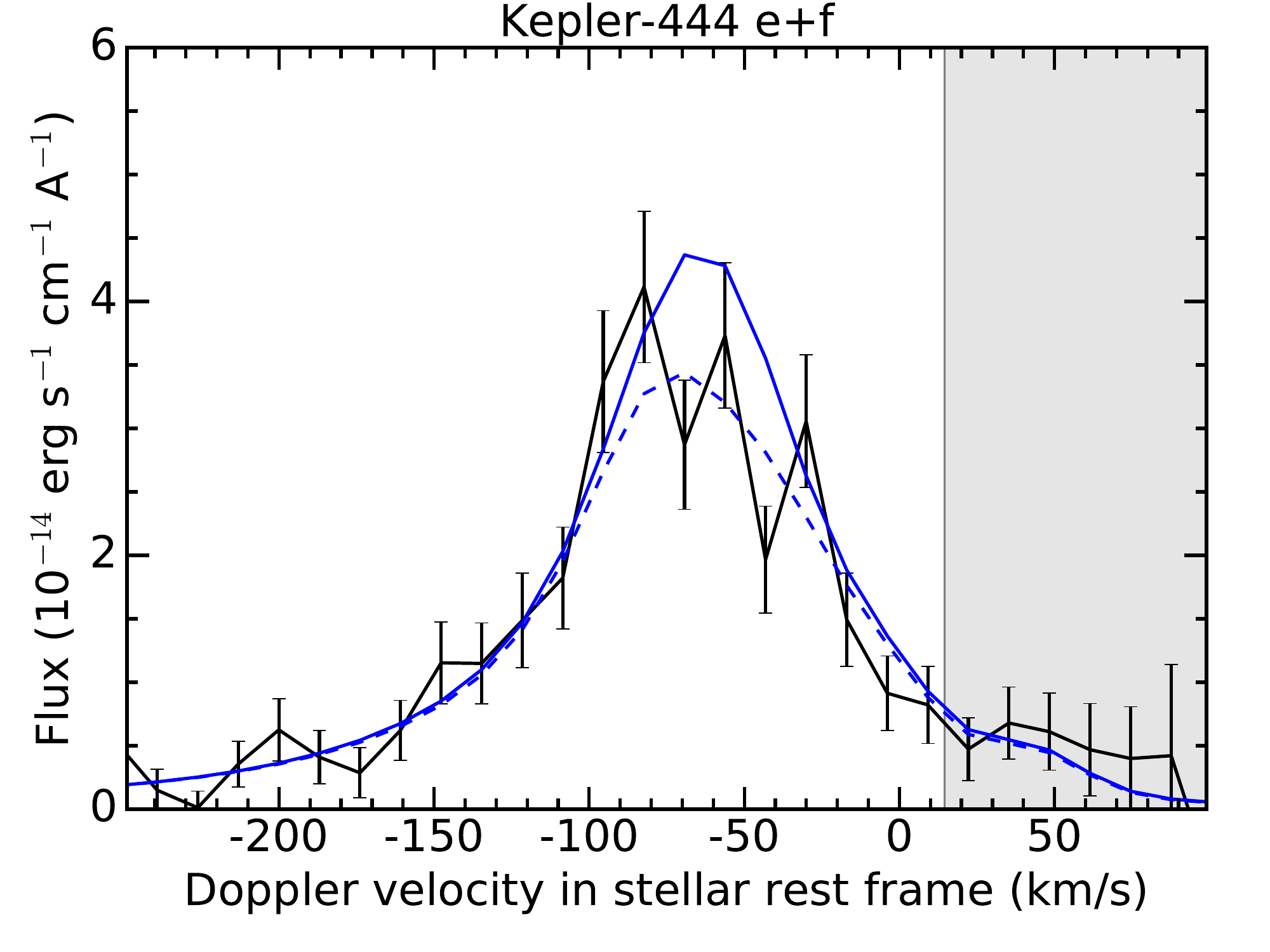}
\includegraphics[trim=2cm 2.2cm 1cm 0cm,clip=true,width=0.317\textwidth]{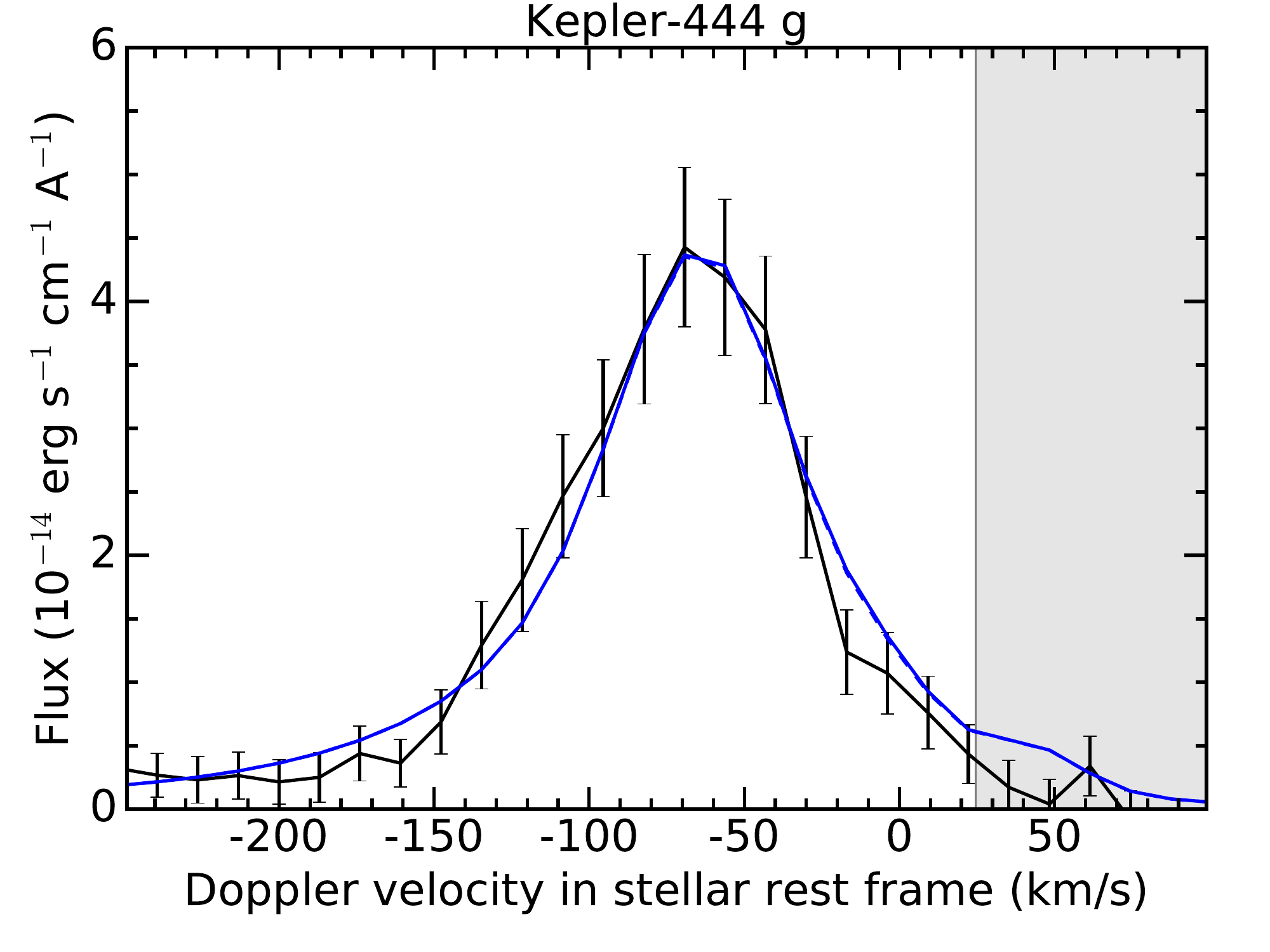}\\
\includegraphics[trim=0cm 2.2cm 1cm 0.7cm,clip=true,width=0.3535\textwidth]{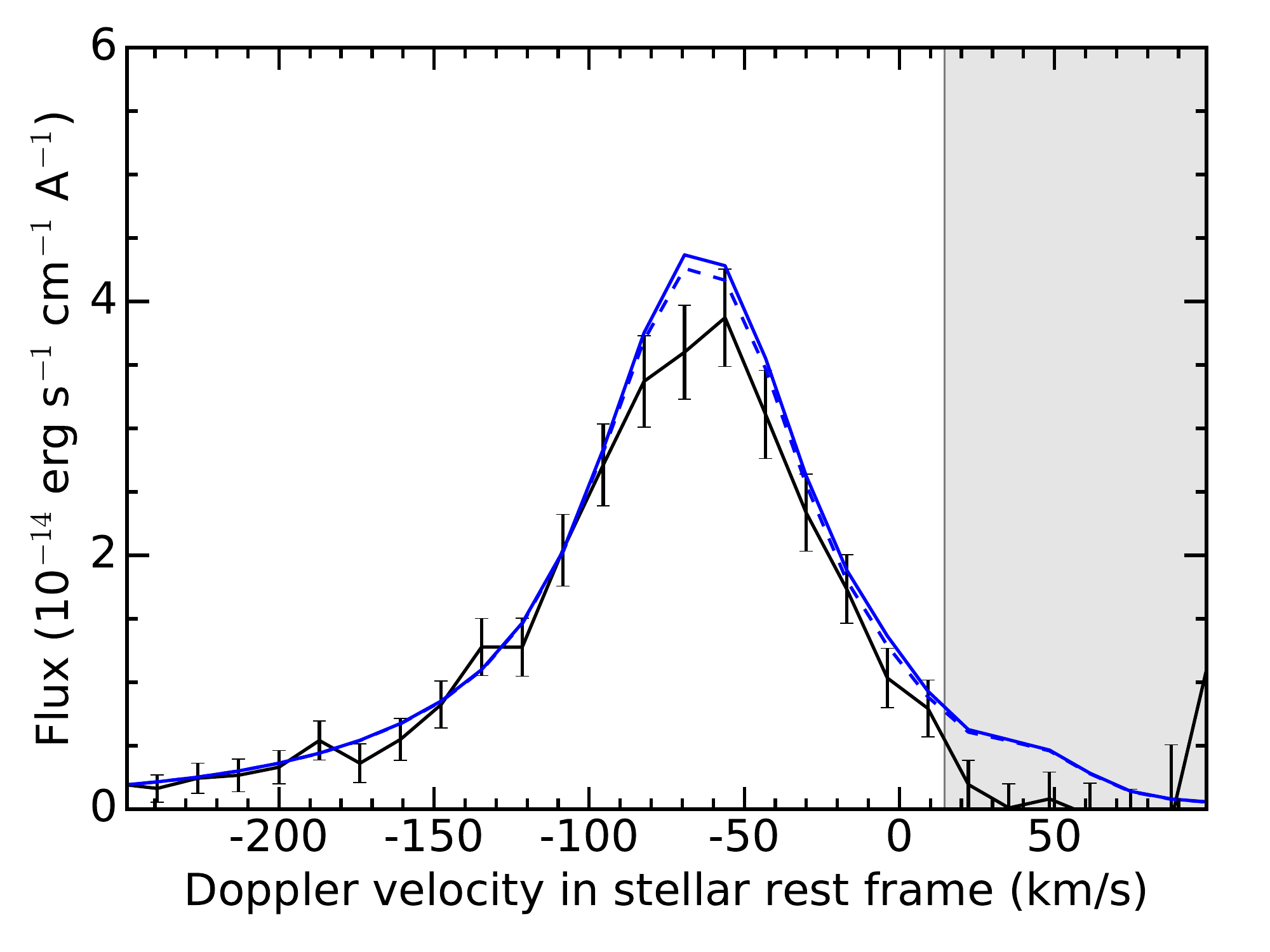}
\includegraphics[trim=2cm 2.2cm 1cm 0.7cm,clip=true,width=0.317\textwidth]{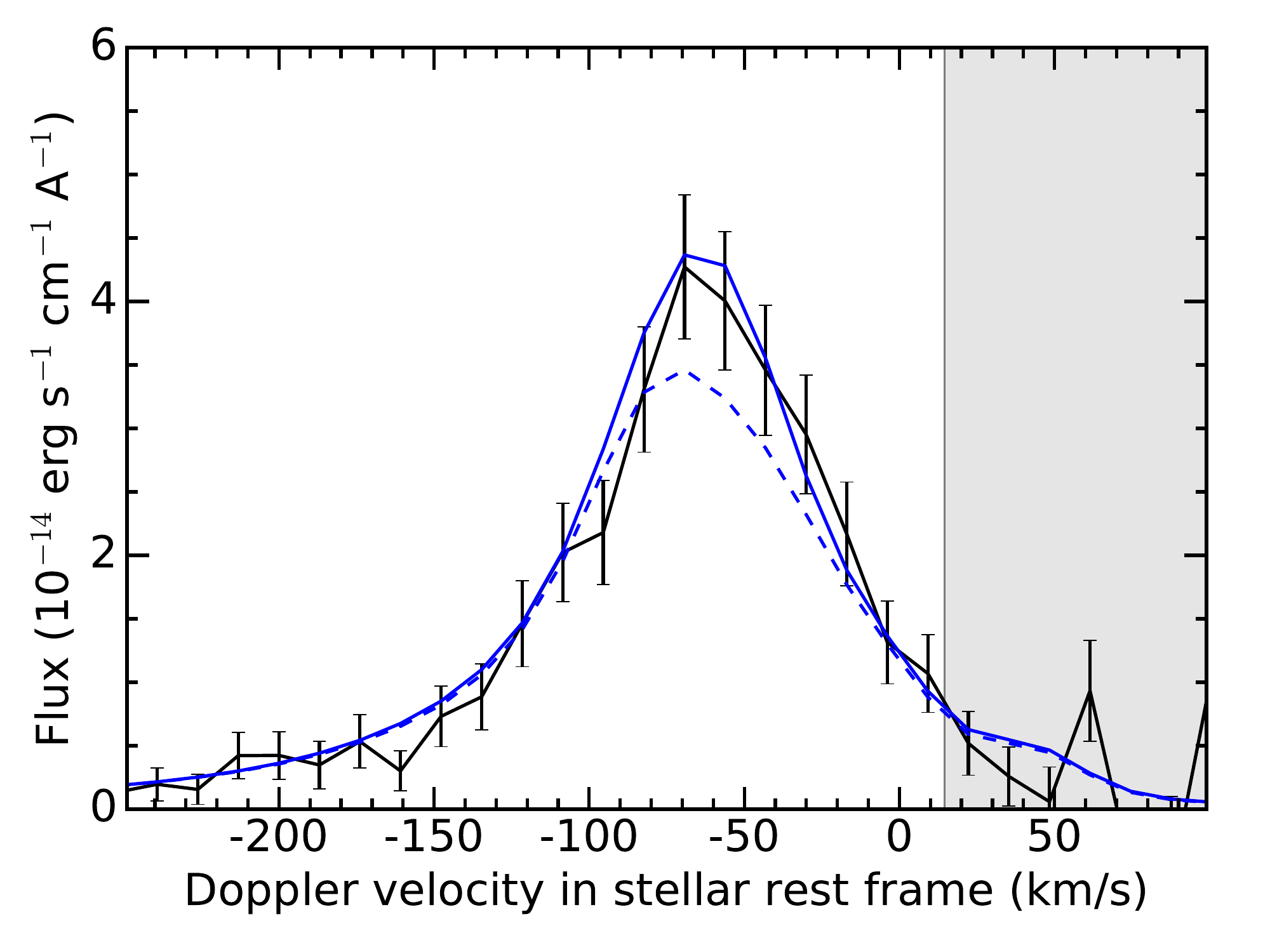}
\includegraphics[trim=2cm 2.2cm 1cm 0.7cm,clip=true,width=0.317\textwidth]{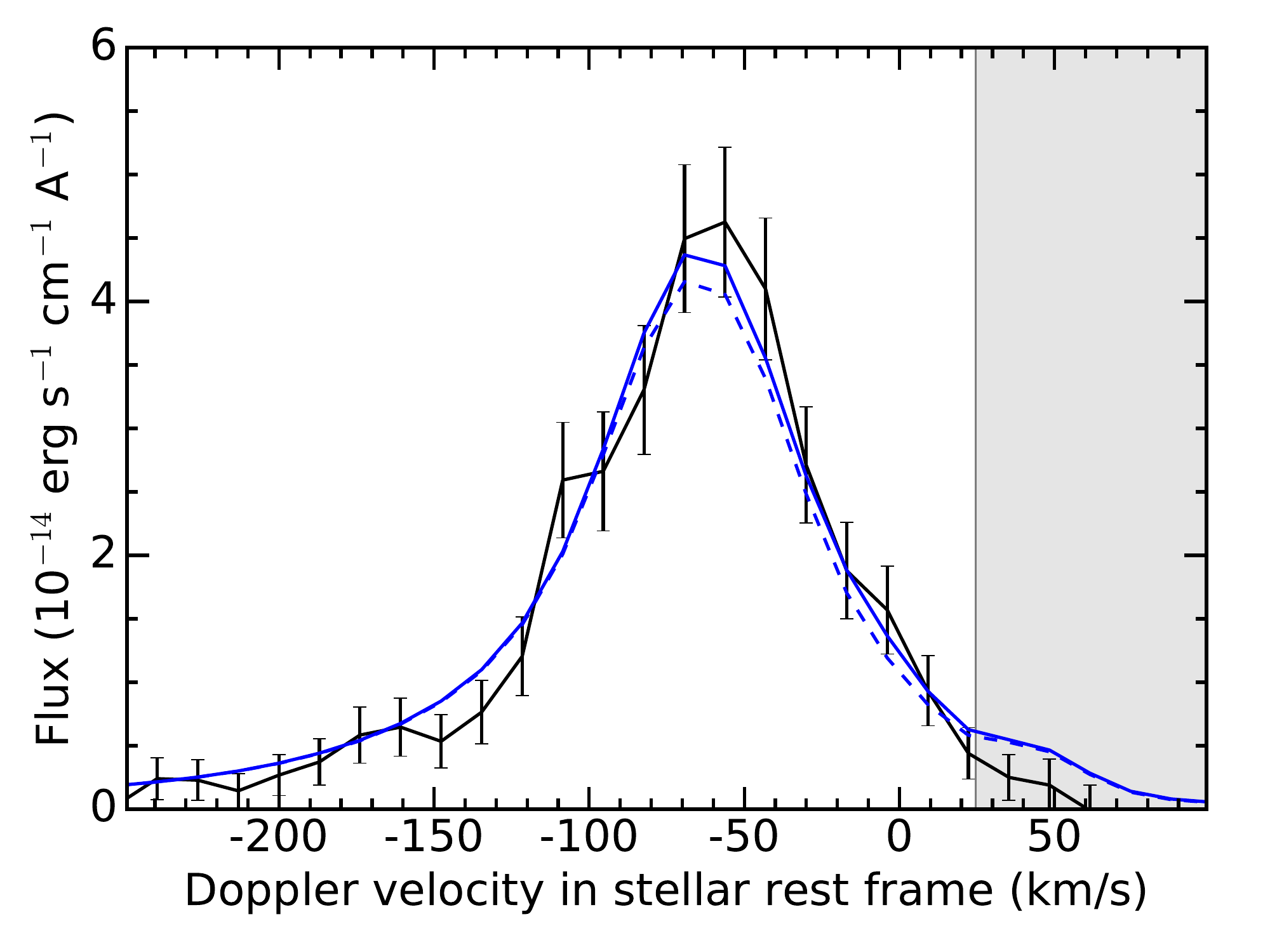}\\
\includegraphics[trim=0cm 2.2cm 1cm 0.7cm,clip=true,width=0.3535\textwidth]{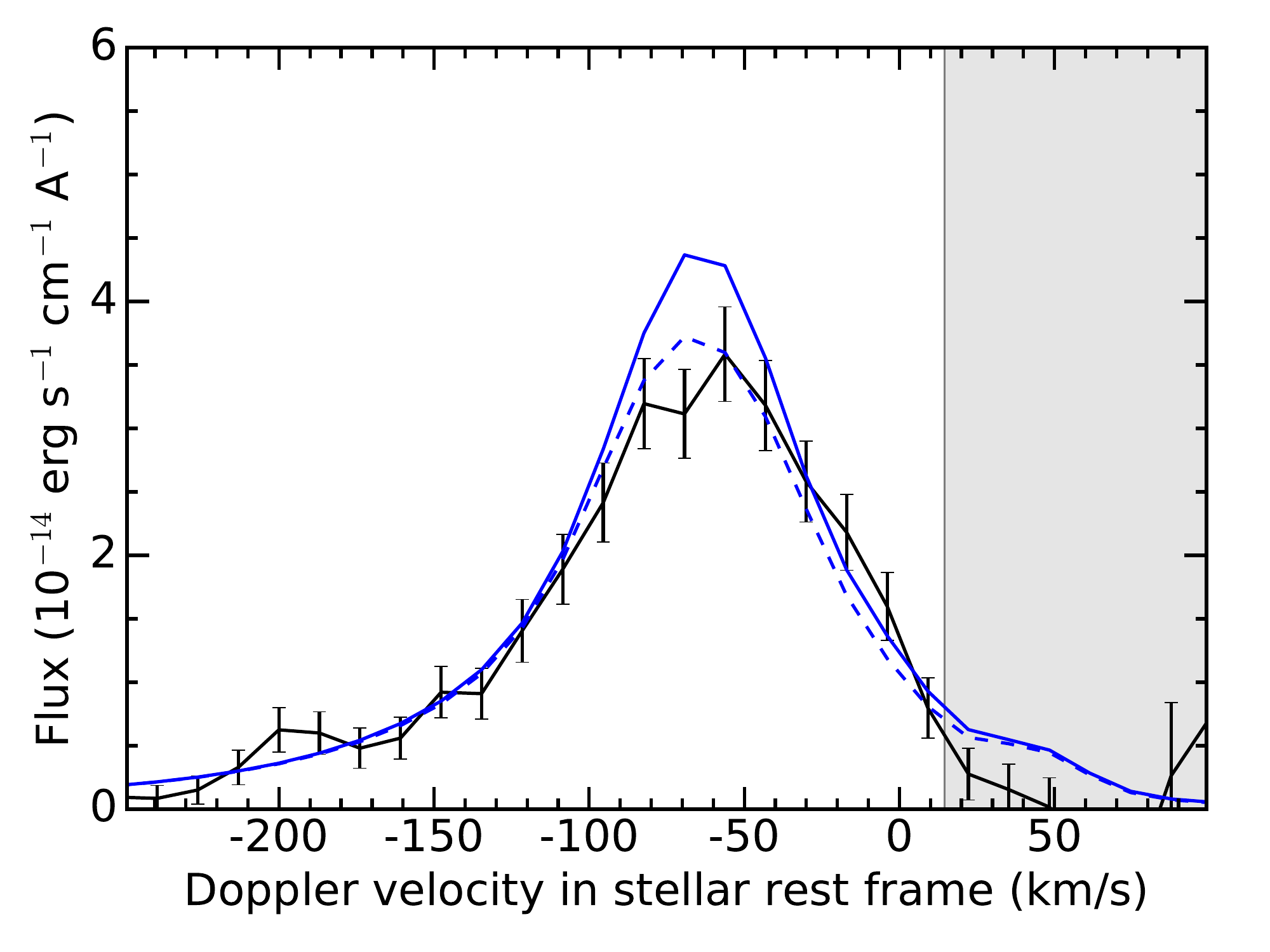}
\includegraphics[trim=2cm 2.2cm 1cm 0.7cm,clip=true,width=0.317\textwidth]{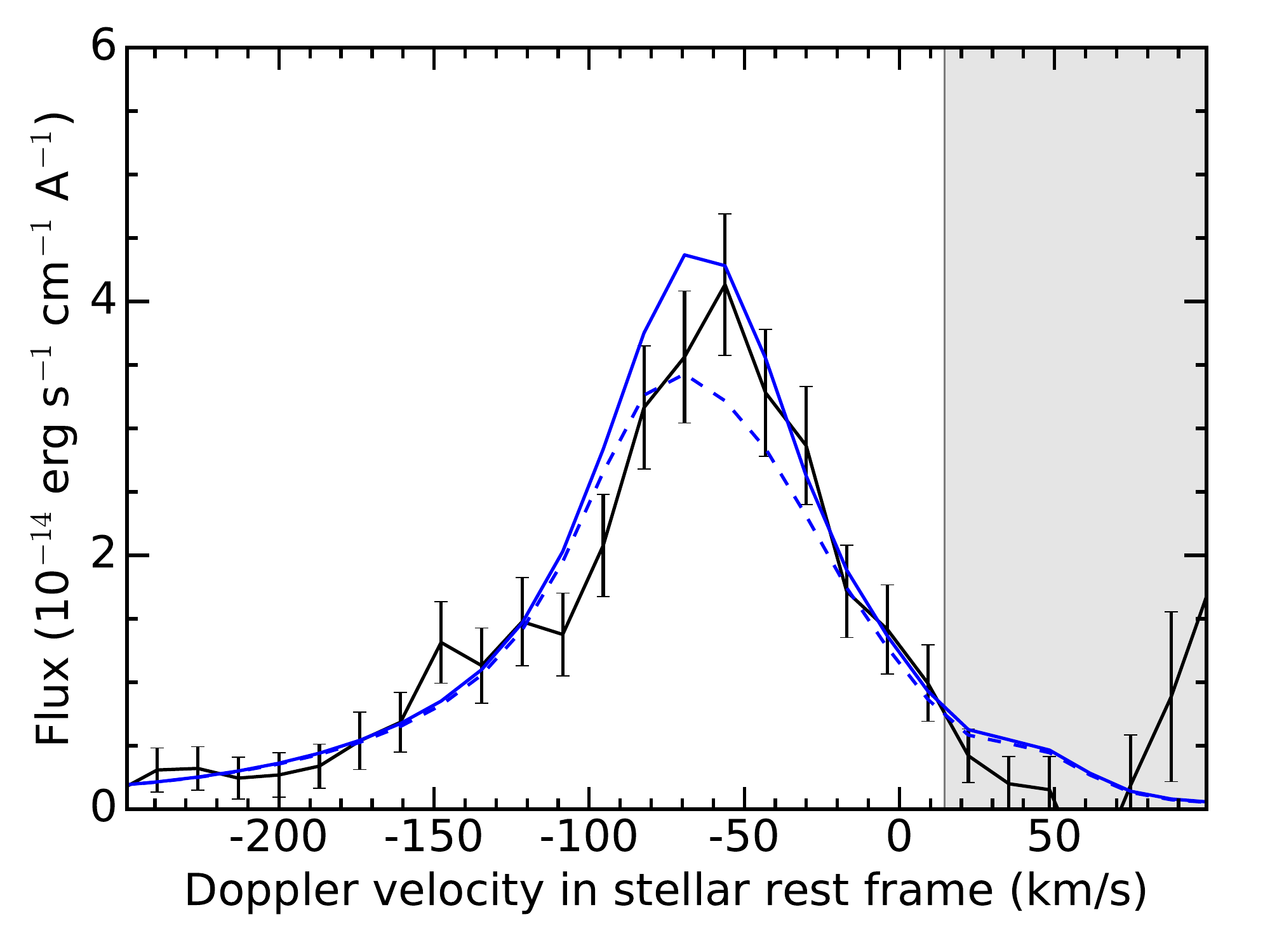}
\includegraphics[trim=2cm 2.2cm 1cm 0.7cm,clip=true,width=0.317\textwidth]{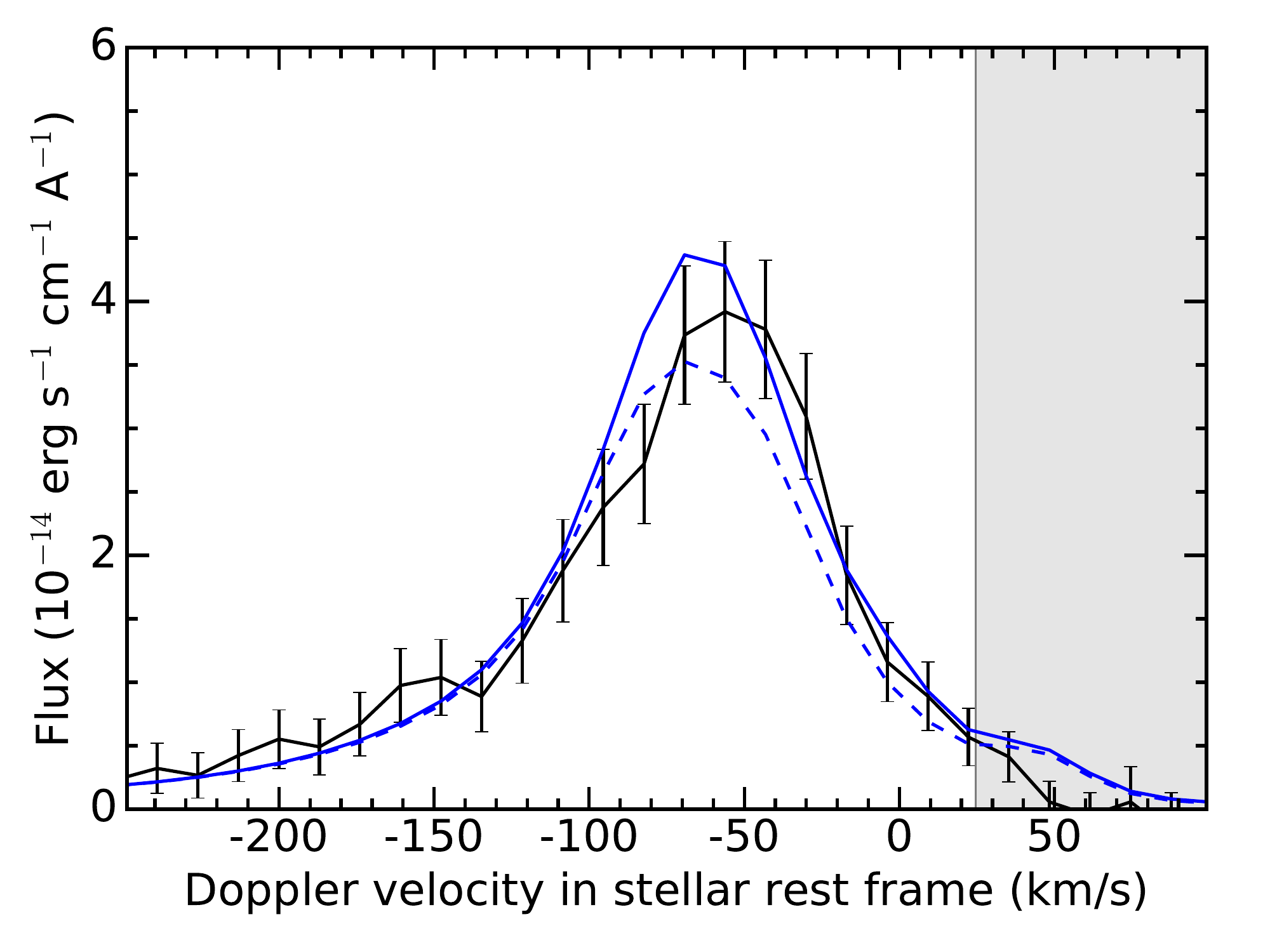}\\
\includegraphics[trim=0cm 2.2cm 1cm 0.7cm,clip=true,width=0.3535\textwidth]{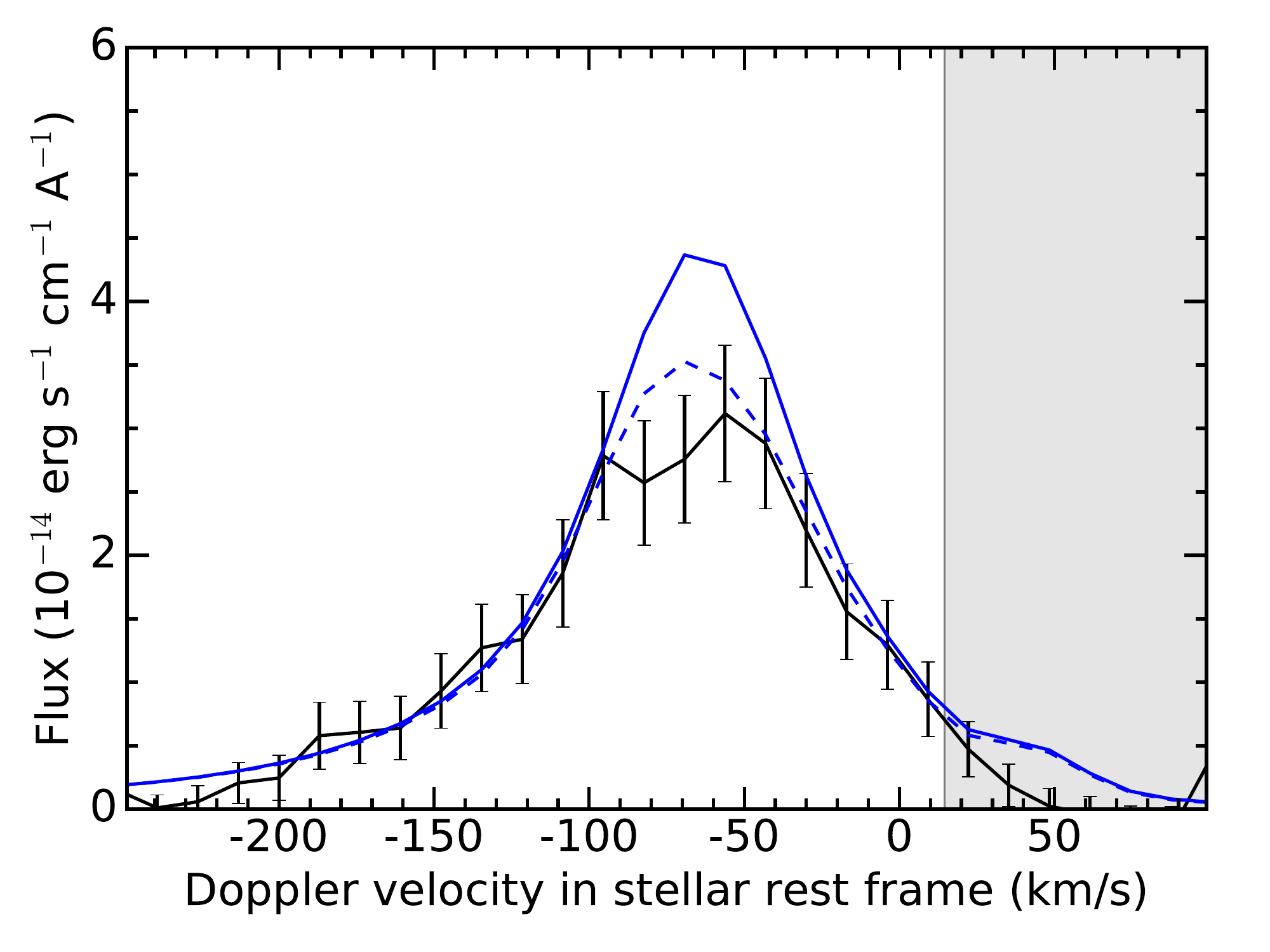}
\includegraphics[trim=2cm 2.2cm 1cm 0.7cm,clip=true,width=0.317\textwidth]{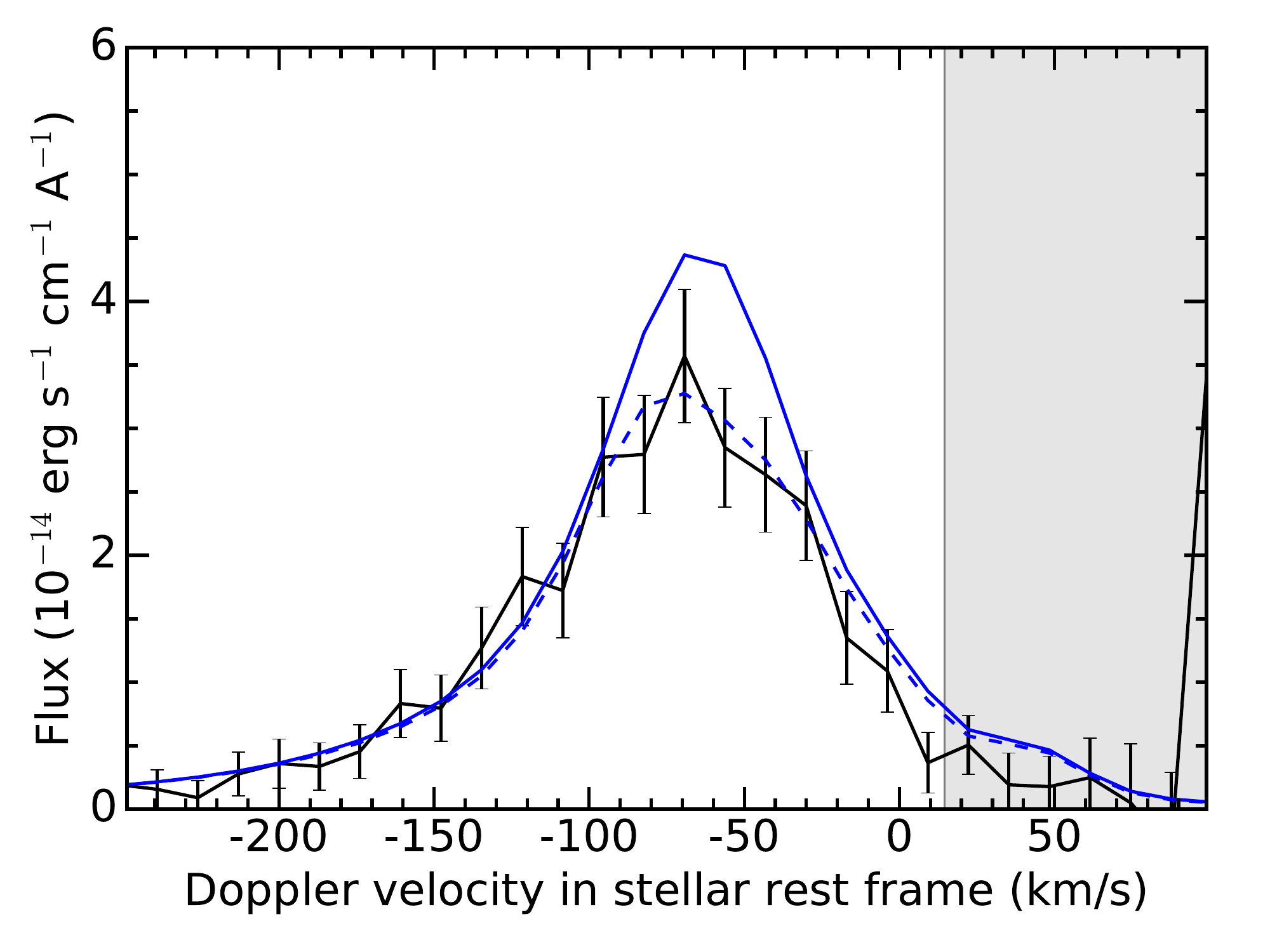}
\includegraphics[trim=2cm 2.2cm 1cm 0.7cm,clip=true,width=0.317\textwidth]{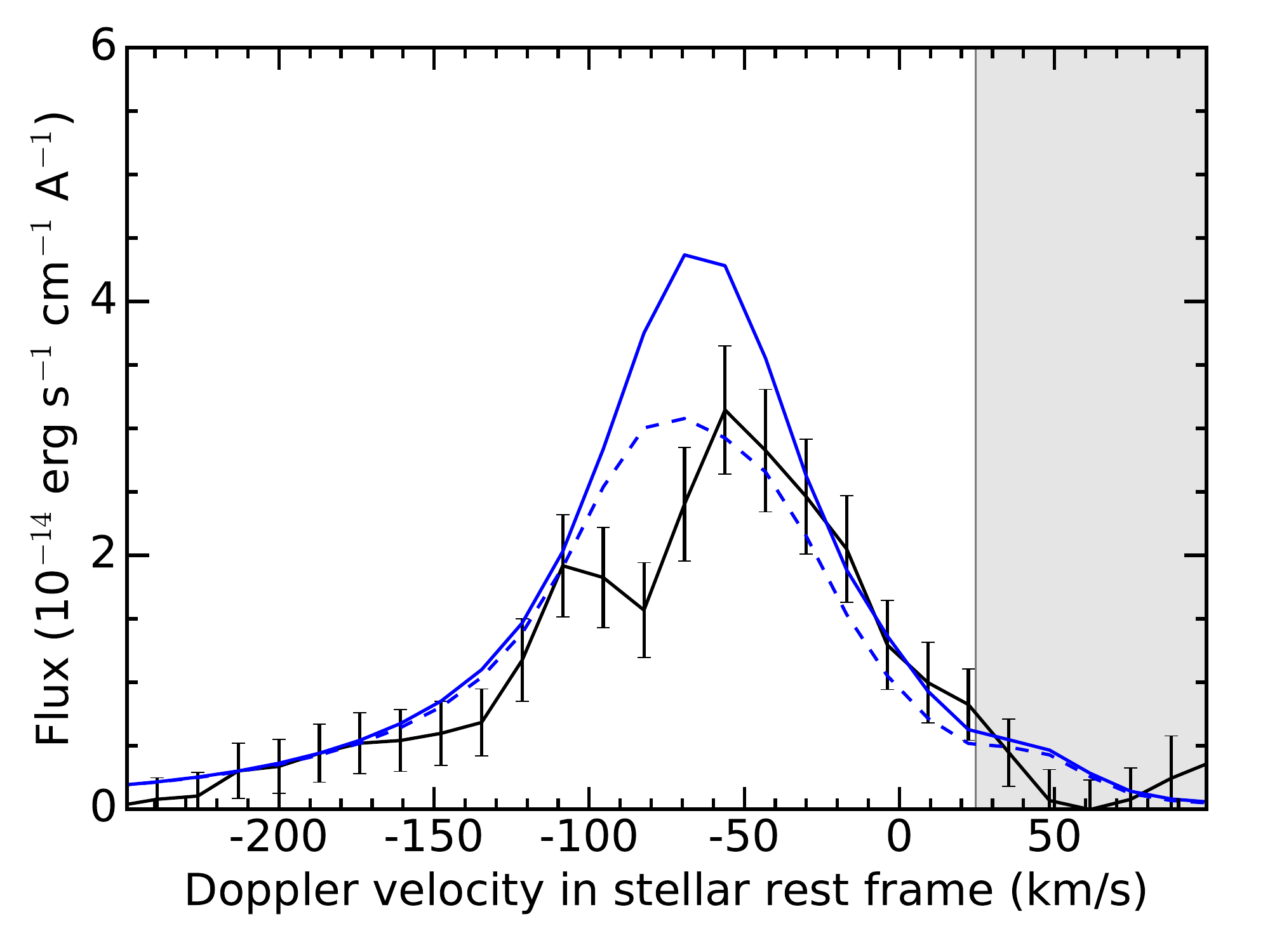}\\
\includegraphics[trim=0cm 0cm 1cm 0.7cm,clip=true,width=0.3535\textwidth]{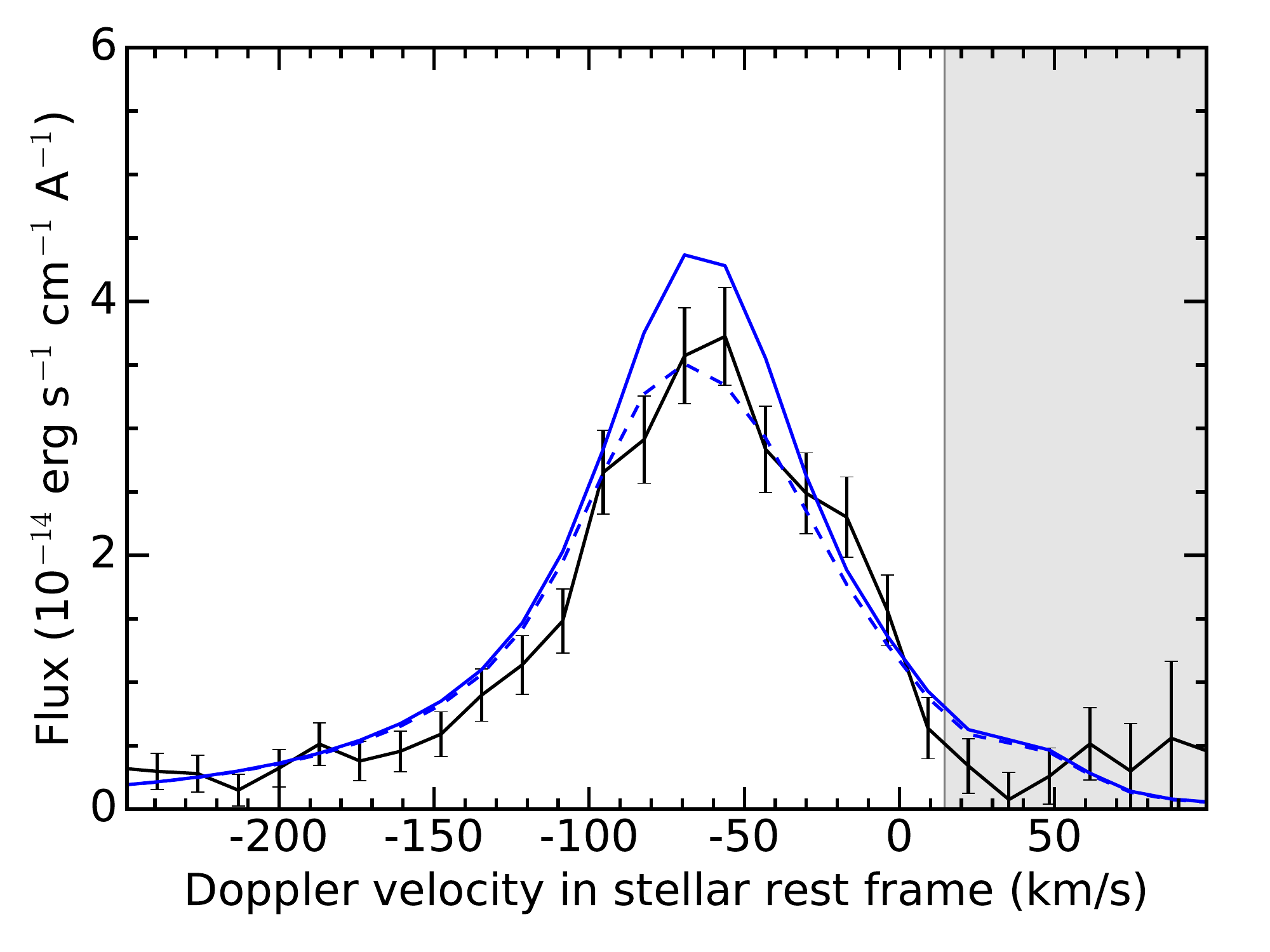}
\includegraphics[trim=2cm 0cm 1cm 0.7cm,clip=true,width=0.317\textwidth]{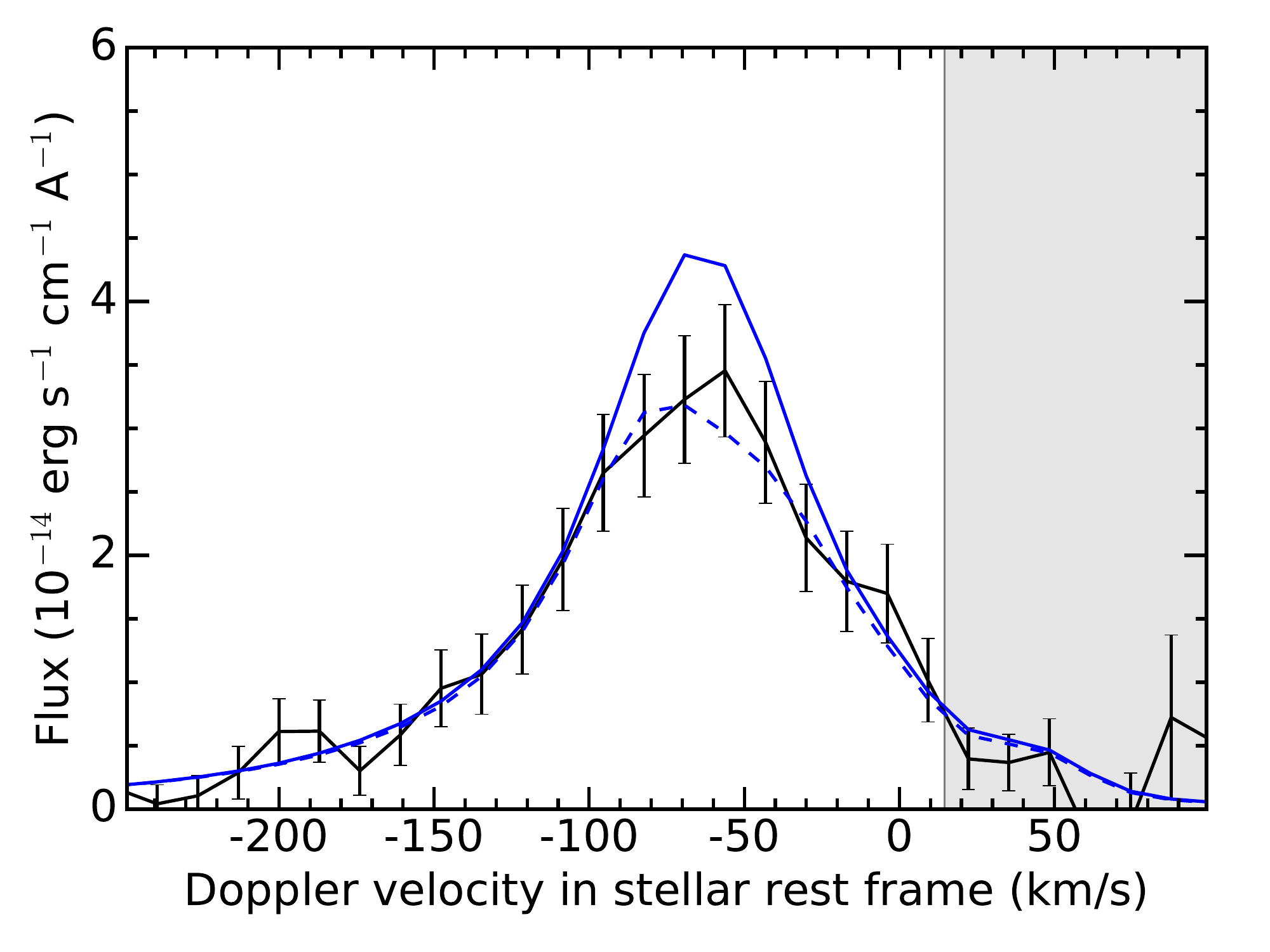}
\includegraphics[trim=2cm -12cm 1cm 13cm,clip=true,width=0.317\textwidth]{Simu2_orb3}\\
\end{minipage}
\caption[]{Kepler-444 spectra associated to the light-curves in Fig.~\ref{fig:light_curve_plf} and Fig.~\ref{fig:LC_fits}.
The first column corresponds to the transits of Kepler-444f in Visits B, C, and F. The second column corresponds to the transit of Kepler-444e in Visit A. The third column corresponds to the inferior conjunction of the hypothetical Kepler-444g in Visit E. In the first and second row, spectra have been averaged over common phase windows after being phased over Kepler-444f ephemeris. The solid-line blue spectrum is the reconstructed out-of-transit stellar line. Dashed-line blue spectra corresponds to simulations that provide a reasonable fit to the data in this scenario. The shaded gray area was excluded from the fits.}
\label{fig:grid_sp_fits}
\end{figure*}

\end{appendix}

\end{document}